\documentclass[10pt,aps,prd,twocolumn,showpacs,superscriptaddress,eqsecnum,longbibliography,nofootinbib]{revtex4-2}
\usepackage{mathrsfs,amsmath,amsthm,latexsym,amssymb,amsfonts,epsfig,cancel,enumerate,graphicx,subcaption,txfonts,overpic,adjustbox}
\usepackage{multirow}
\usepackage{xcolor}
\definecolor{navy}{RGB}{0,0,150}
\usepackage[colorlinks, linkcolor=blue, citecolor=blue, urlcolor=blue, plainpages=false, pdfstartview=FitH]{hyperref}
\usepackage{appendix}
\allowdisplaybreaks
\newcommand{\YO}{Yunnan Observatories, Chinese Academy of Sciences, Kunming 650216, People's Republic of China}

\newcommand{\GZU}{College of Physics, Guizhou University, Guiyang 550025, China}
\newcommand{\HU}{College of Physics and Science and Technology, Hebei University, Baoding 071002, China}

\begin{document}
	
	\title{Connection Between the Shadow Radius and Quasinormal Frequencies for Black Holes in STVG with Perfect Fluid Dark Matter}
	
	\author{Ziqiang Cai}
	\email{gs.zqcai24@gzu.edu.cn}
	\affiliation{\GZU}
	
	\author{Zhi Li}
	\email{lizhi@ynao.ac.cn}
	\affiliation{\YO}

	\author{Zhenglong Ban}
	\email{zlban123@163.com}
	\affiliation{\HU}
	
	\author{Qi-Qi Liang}
	\email{13985671614@163.com}
	\affiliation{\GZU}
	
	\author{Zheng-Wen Long}
	\thanks{Corresponding author}
	\email{zwlong@gzu.edu.cn}
	\affiliation{\GZU}
	
	\begin{abstract}
	We investigate the connection between black hole shadow and quasinormal mode (QNM) spectra in the context of scalar--tensor--vector gravity (STVG) coupled to perfect fluid dark matter (PFDM), characterized by the MOG parameter $\alpha$ and the dark matter intensity $\lambda$. Employing complementary methods—namely the sixth-order WKB approximation, Pad\'e resummation, and time-domain numerical integration—we compute QNM frequencies for scalar ($s=0$), electromagnetic ($s=1$), and axial gravitational ($s=2$) perturbations. Both the real part of the QNM frequencies and the peak height of the effective potential display a consistent parametric dependence: they increase with $\lambda$ yet decrease with growing $\alpha$. In the eikonal limit ($l \gg 1$), we derive an exact analytical link between the shadow radius $R_{\rm sh}$ and the QNM frequency $\omega_R$. Noting that $R_{\rm sh}$ is determined by the critical impact parameter $b_c = r_{\rm ph}/\sqrt{f(r_{\rm ph})}$, while $\omega_R = \Omega l$ with photon angular velocity $\Omega = \sqrt{f(r_{\rm ph})}/r_{\rm ph}$, we obtain the precise relation $\omega_R = l / b_c$, identifying $R_{\rm sh} \equiv b_c$ for an asymptotically flat observer. This prediction is robustly validated by numerical results across all three computational approaches at large multipole numbers. Our findings reveal that the black hole shadow and gravitational ringdown are not independent phenomena, but dual observational signatures of the same underlying structure—the unstable photon orbit—thereby offering a unified multi-messenger framework to simultaneously constrain modified gravity and dark matter in the strong-field regime.
	\end{abstract}
	
	\maketitle
	\section{Introduction}
	Among the fundamental theories of modern physics, General Relativity (GR) occupies a privileged position as the standard framework for classical gravity. Its predictions have been confirmed with remarkable precision across diverse astrophysical and cosmological scales \cite{Will:2014kxa,LIGOScientific:2016aoc}, and its mathematical structure remains influential even in contexts traditionally associated with particle physics. Even though GR has been exquisitely confirmed by a host of observations—including solar system tests, binary pulsar dynamics, gravitational wave signals, and black hole shadow measurements—it fails to reconcile with quantum mechanics and exhibits singularities, signaling its incompleteness \cite{Moffat:2005si,Tsupko:2009zz}. Motivated by the persistent challenges facing GR, numerous modified gravity theories have emerged. Among them, STVG, or MOG \cite{Nishonov:2025pgq,Nishonov:2024uum}, stands out for its ability to address key astrophysical anomalies. By augmenting the Einstein–Hilbert action with a scalar field—through which Newton's constant becomes spacetime-dependent—and a massive vector field mediating a repulsive interaction, STVG provides a compelling dark-matter-free explanation for galaxy rotation curves, cluster kinematics, and lensing observations \cite{Murodov:2023one}. Complementing efforts in modified gravity, the PFDM paradigm treats dark matter as a barotropic continuum. Embedding PFDM into black hole geometries induces nontrivial modifications to spacetime, with observable consequences for orbital dynamics, accretion flows, and thermodynamic stability \cite{Khan:2023nul}. The primary goal of this work is to explore how the interplay between STVG and PFDM manifests in black hole shadows and quasinormal ringing. This investigation is motivated by the possibility that such a combined framework could reproduce or reinterpret astrophysical phenomena without invoking conventional particle dark matter.
	
	Moreover, the successful imaging of black hole shadows by the Event Horizon Telescope (EHT) has established shadow size as a robust observable for probing strong-field gravity. This development has motivated numerous studies of shadow radii in modified gravity theories and non-vacuum spacetimes \cite{Okyay:2021nnh,Roy:2021uye,Khodadi:2020jij,Vagnozzi:2022moj,Wang:2018prk,Cunha:2019hzj,Pantig:2020uhp}, where deviations from GR—or the presence of exotic matter—can imprint measurable signatures. In the context of STVG coupled with PFDM, we focus specifically on how the shadow radius varies with the MOG parameter $\alpha$ and the PFDM parameter $\lambda$, providing a direct link between theoretical parameters and an EHT-accessible quantity.
	
	Because black holes relax to equilibrium through damped oscillations characterized by QNMs, these complex frequencies serve as direct probes of strong-field gravity. The real component reflects the mode's oscillation frequency, while the imaginary part quantifies its decay \cite{Konoplya:2011qq,Berti:2009kk}. Mathematically, QNMs emerge as eigenvalues of a Schr\"odinger-like wave equation under dissipative boundary conditions—ingoing at the horizon and outgoing at infinity \cite{Chandrasekhar:1985kt,Leaver:1985ax}. In practice, their computation relies on semi-analytical or numerical techniques such as the WKB method, Pad\'e approximants, continued fractions, and time-domain integration, given the intractability of exact solutions for generic potentials \cite{Nollert:1999ji,Matyjasek:2019eeu,Cardoso:2019rvt}. Originally introduced in Refs. \cite{Schutz:1985km,Iyer:1986nq,Iyer:1986np} to compute QNMs of the Schwarzschild black hole, the WKB method has proven to be a simple yet powerful semi-analytical approach for investigating black hole perturbations. Subsequent applications extended this framework to rotating (Kerr) and charged (Reissner–Nordström) spacetimes \cite{Kokkotas:1988fm,Kokkotas:1991vz,Konoplya:2003ii,Zhidenko:2003wq}. Since then, a substantial body of research has explored QNMs in a wide range of black hole geometries and matter environments, including those arising in modified gravity and exotic compact objects \cite{Chan:1996yk,Horowitz:1999jd,Hod:1998vk,Chen:2011dc,Stefanov:2010xz,Konoplya:2020jgt,Konoplya:2020fwg,Fernandes:2021qvr,Siqueira:2022tbc,Torres:2020tzs,Okyay:2021nnh,Oshita:2018fqu,Jha:2023rem,Barman:2024hwd}. Although QNMs and the shadow radius seem unrelated, they are connected in the eikonal limit: the real part of the QNM frequency corresponds to the orbital frequency of the photon sphere \cite{Cardoso:2008bp,Hod:2013jhd,Wei:2019jve}, and this link extends to strong-lensing phenomena that define the shadow boundary \cite{Stefanov:2010xz}. Building on this insight, we calculate QNMs via WKB, Pad\'e, and time-domain methods and explicitly examine their correlation with the shadow radius for black holes in STVG surrounded by PFDM. It should be noted that the correspondence between eikonal quasinormal modes and unstable circular null geodesics is not universally valid for all gravitational theories and types of perturbations. As pointed out in Refs. \cite{Konoplya:2017wot,Konoplya:2022gjp}, the duality may break down for gravitational perturbations in higher-curvature gravity theories and does not cover the full quasinormal mode spectrum, such as the purely imaginary non-oscillatory modes in asymptotically de Sitter spacetimes. Nevertheless, the correspondence remains rigorous for test fields (scalar, electromagnetic) and axial gravitational perturbations in asymptotically flat spacetimes with a well-behaved single-peak effective potential, which is exactly the case considered in this work and has been analytically verified for general spherically symmetric black holes in Ref. \cite{Cuadros-Melgar:2020kqn}.
	
	The structure of this paper is as follows. The spacetime geometry of STVG black holes embedded in PFDM is reviewed in Sec. \ref{spacetime}. The dependence of the effective potential on the model parameters is examined, the connection between the number of photon orbits and the impact parameter is explored, and integrated images of null geodesics are presented. Sec. \ref{qnms} is devoted to the computation of QNMs for scalar, electromagnetic, and axial gravitational perturbations, employing the WKB approximation, Pad\'e approximants, and time-domain integration. Moreover, within the eikonal limit, we derive an analytical relation between the real part of the quasinormal frequencies and the black hole shadow radius using the WKB method. We summarize our main results and discuss their implications in Sec. \ref{conclusion}.
	\section{Black Hole Spacetime in STVG with Perfect Fluid Dark Matter: Effective Potential and Shadow}
	\label{spacetime}
\begin{figure*}[htbp]
		\centering
		\begin{subfigure}{0.38\textwidth}
			\includegraphics[width=\linewidth]{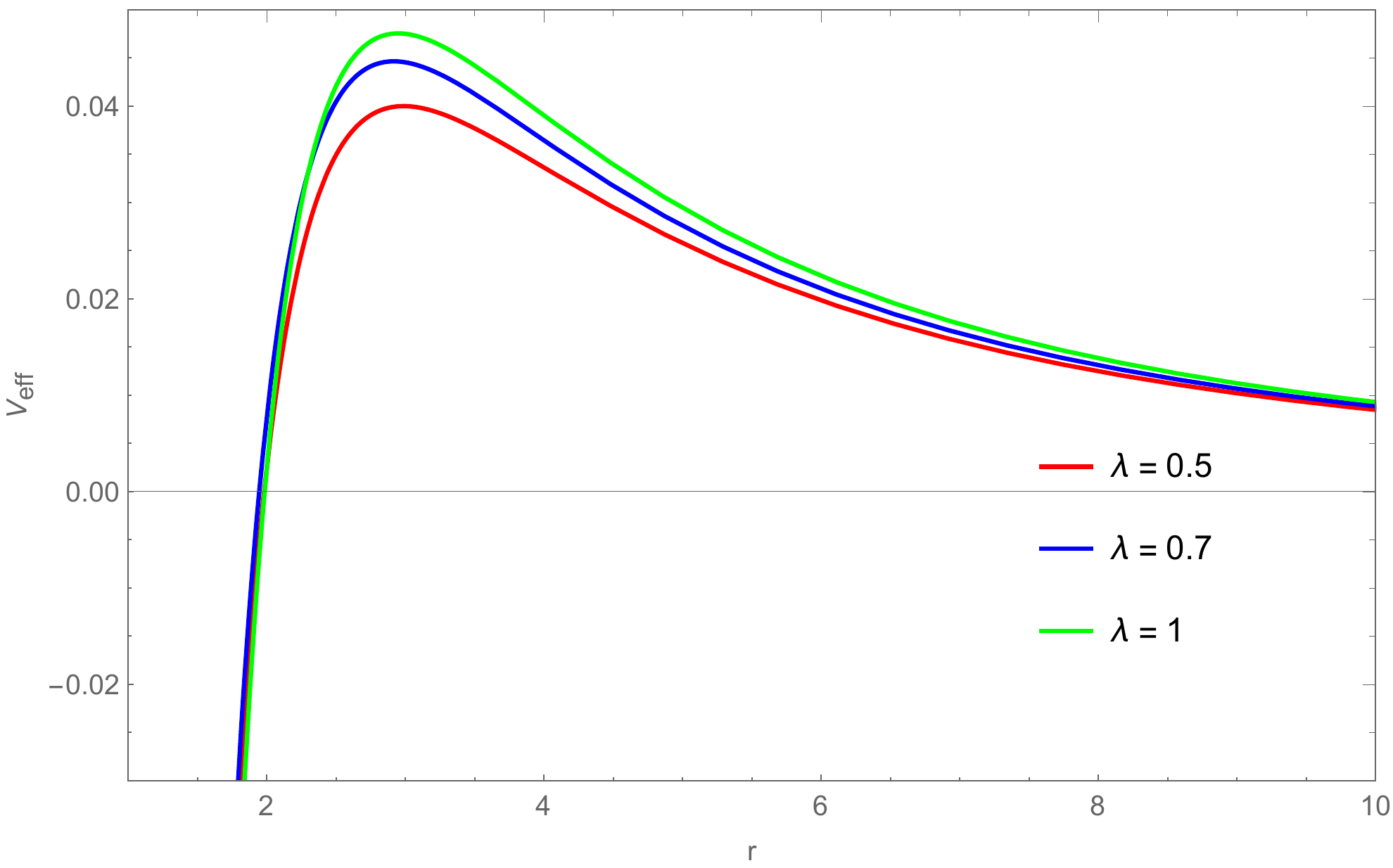}
		\end{subfigure}%
		\hspace{0.13\textwidth}%
		\begin{subfigure}{0.38\textwidth}
			\includegraphics[width=\linewidth]{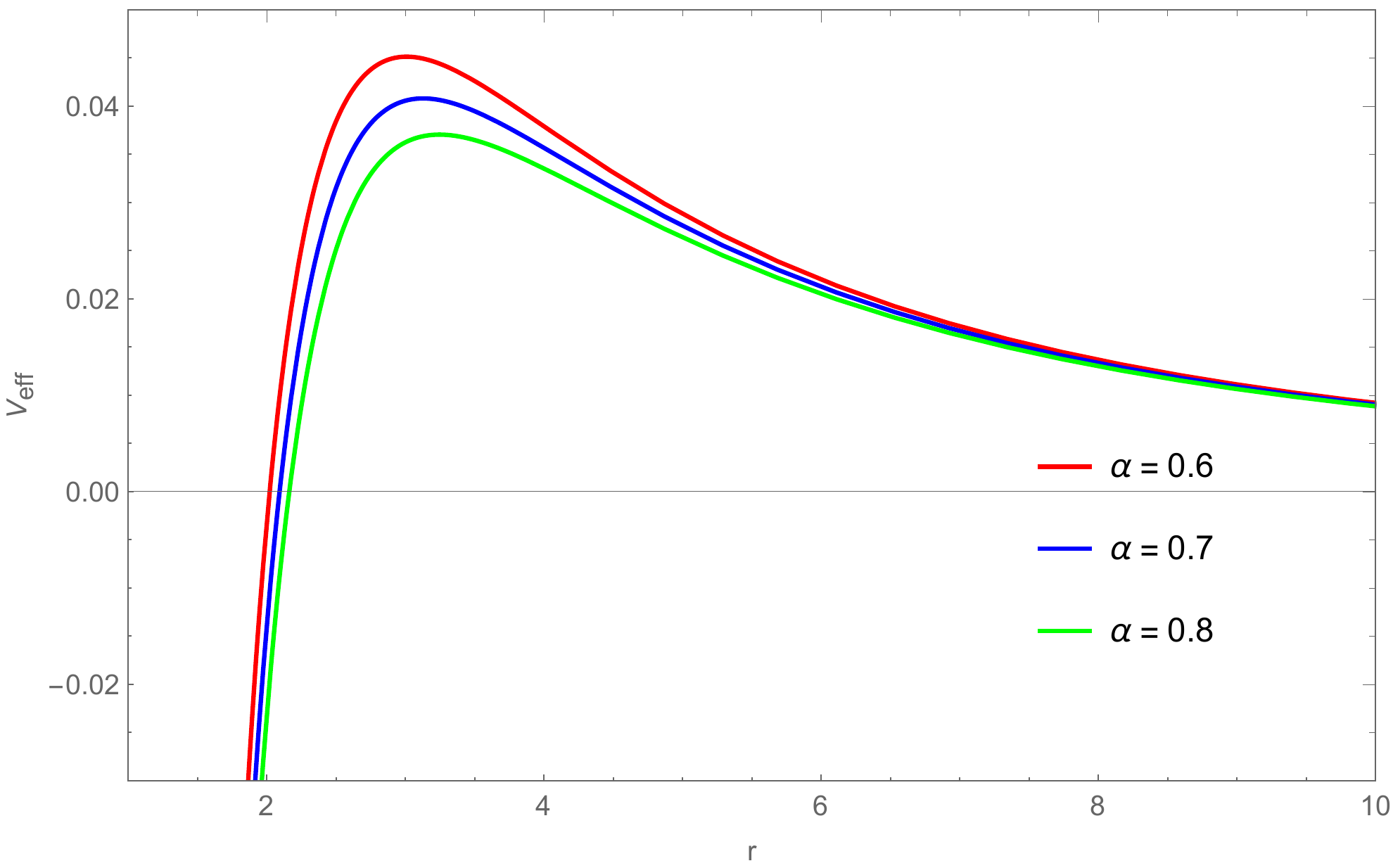}
		\end{subfigure}
		
		\caption{Effective potential curves for black holes under different parameter values: $\alpha = 0.55$ (left) and $\lambda = 1$ (right).}
		\label{figVeff}
	\end{figure*}

	We consider a static, spherically symmetric black hole (BH) in STVG (also known as modified gravity, MOG) surrounded by PFDM. The PFDM is characterized by a barotropic equation of state $P=\frac{1}{2}\rho$ and a density distribution $\rho = -\frac{\lambda}{8\pi r^{3}}$, where$\lambda$ denotes the PFDM parameter that quantifies the contribution of dark matter to the spacetime geometry. The spacetime geometry of this STVG-PFDM black hole system is described by the line element:
	\begin{equation}
		ds^{2}=-f(r)dt^{2}+\frac{dr^{2}}{f(r)}+r^{2}\left(d\theta^{2}+\sin^{2}\theta d \phi^{2}\right),\label{xianyuan}
	\end{equation}
	where the metric function $f(r)$ is derived from solving the Einstein field equations in the STVG framework with PFDM contributions \cite{Nishonov:2025pgq}:
	\begin{equation}
		f(r)=1-\frac{2(1+\alpha)M}{r}+\frac{\alpha(1+\alpha)M^2}{r^2}+\frac{\lambda}{r}\ln\frac{r}{|\lambda|}.\label{fr}
	\end{equation}
	Geodesic motion of a test particle in curved spacetime follows from the Euler–Lagrange equations applied to the Lagrangian
	\begin{equation}
		\mathcal{L} = \frac{1}{2} g_{\mu\nu} \dot{x}^\mu \dot{x}^\nu,
		\label{Lag}
	\end{equation}
	where an overdot denotes differentiation with respect to an affine parameter $\tau$. For a static, spherically symmetric metric, the existence of timelike and rotational Killing vectors ensures conservation of energy $E$ and angular momentum $L$. Exploiting spherical symmetry, we set $\theta = \pi/2$ (so that $\dot{\theta} = 0$) and obtain
	\begin{equation}
		E = f(r)\,\dot{t}, \qquad L = r^2\,\dot{\phi}.
		\label{EL}
	\end{equation}
	For null geodesics, the condition $\mathcal{L} = 0$ combined with Eqs. (\ref{Lag}) and (\ref{EL}) yields the radial equation of motion:
	\begin{equation}
		\left(\frac{dr}{d\phi}\right)^{2}=r^{4}\left(\frac{1}{b^{2}}-V_{\mathrm{eff}}\right),\label{guiji}
	\end{equation}
	with $b \equiv L/E$ the impact parameter. The effective potential,
	\begin{equation}
		V_{\mathrm{eff}}(r) = \frac{f(r)}{r^2},
		\label{eqVeff}
	\end{equation}
	encodes the influence of spacetime curvature on photon propagation and plays a central role in determining the black hole shadow. To develop a more intuitive grasp of the underlying spacetime geometry of a black hole in STVG surrounded by PFDM, we plot the effective potential $V_{\mathrm{eff}}(r)$ as a function of the radial coordinate $r$ in Fig. \ref{figVeff}. As shown in Fig. \ref{figVeff}, the peak of the effective potential decreases with increasing STVG parameter $\alpha$, while it increases with the PFDM parameter $\lambda$.
	
\begin{figure}[htbp]
	\centering
	\begin{subfigure}{0.4\textwidth}
		\includegraphics[width=2.6in, height=3.5in, keepaspectratio]{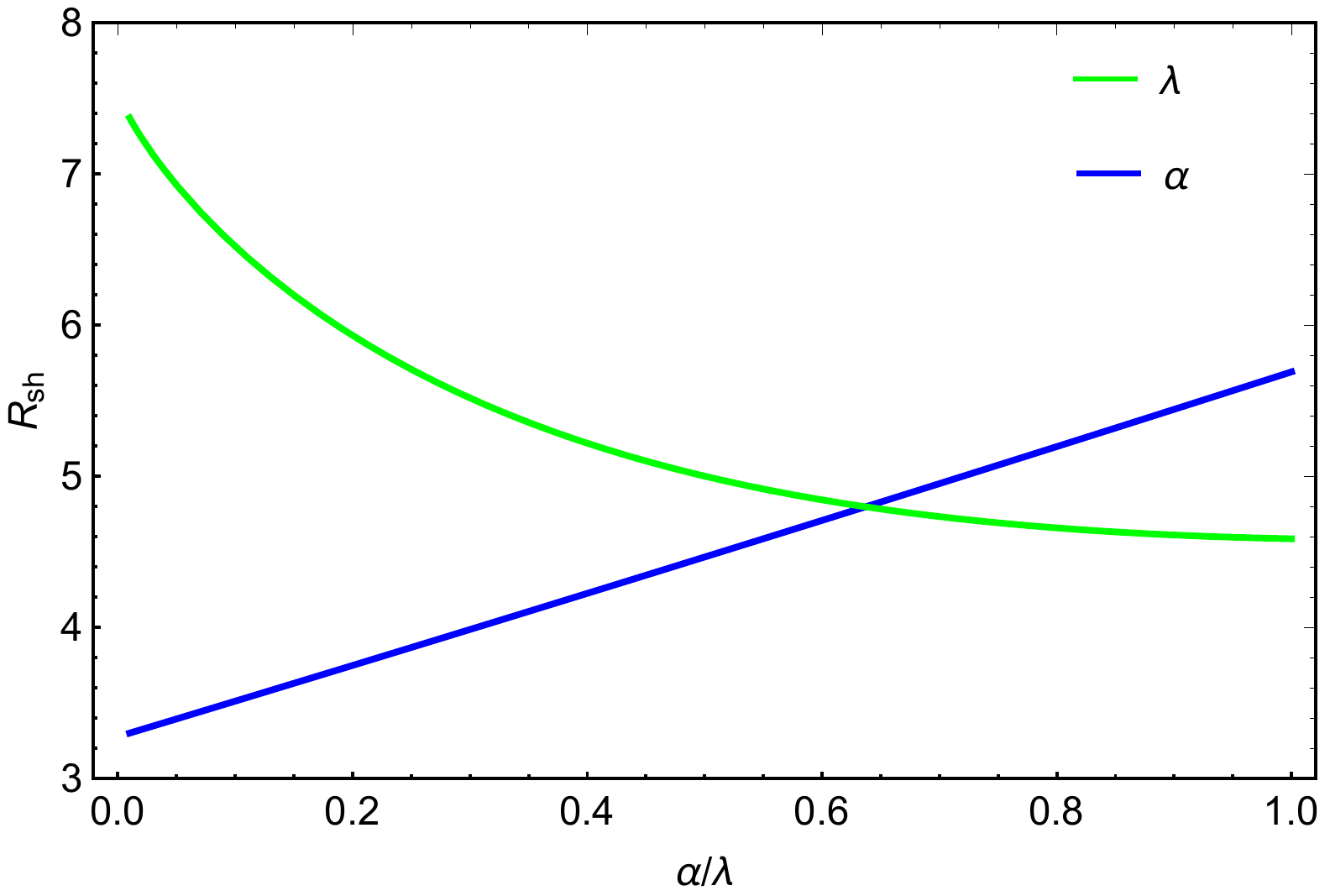}
	\end{subfigure}
	\caption{Variation of the shadow radius $R_{sh}$ with respect to parameters $\alpha$ and $\lambda$.}
	\label{rsh}
    \end{figure}

     The unstable photon sphere plays a pivotal role in shaping the black hole shadow. It corresponds to the local maximum of the effective potential $V_{\mathrm{eff}}(r)$, satisfying
     \begin{equation}
	\left. \frac{dV_{\mathrm{eff}}}{dr} \right|_{r = r_{\mathrm{ph}}} = 0,
	\quad
	\left. \frac{d^2 V_{\mathrm{eff}}}{dr^2} \right|_{r = r_{\mathrm{ph}}} < 0.
	\label{rph}
    \end{equation}
    This orbit separates photons that are captured by the black hole from those that escape to infinity. The radius of the photon sphere, $r_{\mathrm{ph}}$, directly sets the scale of the black hole shadow as seen by a distant static observer. The critical impact parameter $b_c$ is defined through the condition for unstable circular photon orbits and is given by
    \begin{equation}
    	b_c = \frac{r_{\mathrm{ph}}}{\sqrt{f(r_{\mathrm{ph}})}}.
    	\label{Rs}
    \end{equation}
	In Fig. \ref{rsh}, we illustrate the variation of the shadow radius $R_{sh}=b_c$ with respect to parameters $\alpha$ and $\lambda$. It is evident from the figure that $R_{sh}$ increases as $\alpha$ increases, while it decreases as $\lambda$ increases.
	
    The trajectory of photons around a black hole in STVG with PFDM is analyzed by employing the substitution $u = 1/r$. The resulting orbital equation is expressed as
    \begin{equation}
    	\left( \frac{du}{d\phi} \right)^2 = \frac{1}{b^2} - u^2 f\!\left( \frac{1}{u} \right) \equiv G(u).\label{guiji2}
    \end{equation}
    The critical impact parameter $b_c$ separates photon trajectories into two classes: those that are captured ($b < b_c$) and those that are deflected ($b > b_c$). The corresponding azimuthal angles are computed via
    \begin{align}
    	\phi &= \int_{0}^{u_h} \frac{du}{\sqrt{G(u)}} \quad (b < b_c), \\
    	\phi &= 2 \int_{0}^{u_m} \frac{du}{\sqrt{G(u)}} \quad (b > b_c),
    \end{align}
    where $u_h = 1/r_h$ and $u_m$ is the outermost root of $G(u) = 0$.
    
\begin{figure}[htbp]
    	\centering
    	\begin{subfigure}{0.4\textwidth}
    		\includegraphics[width=2.6in, height=3.5in, keepaspectratio]{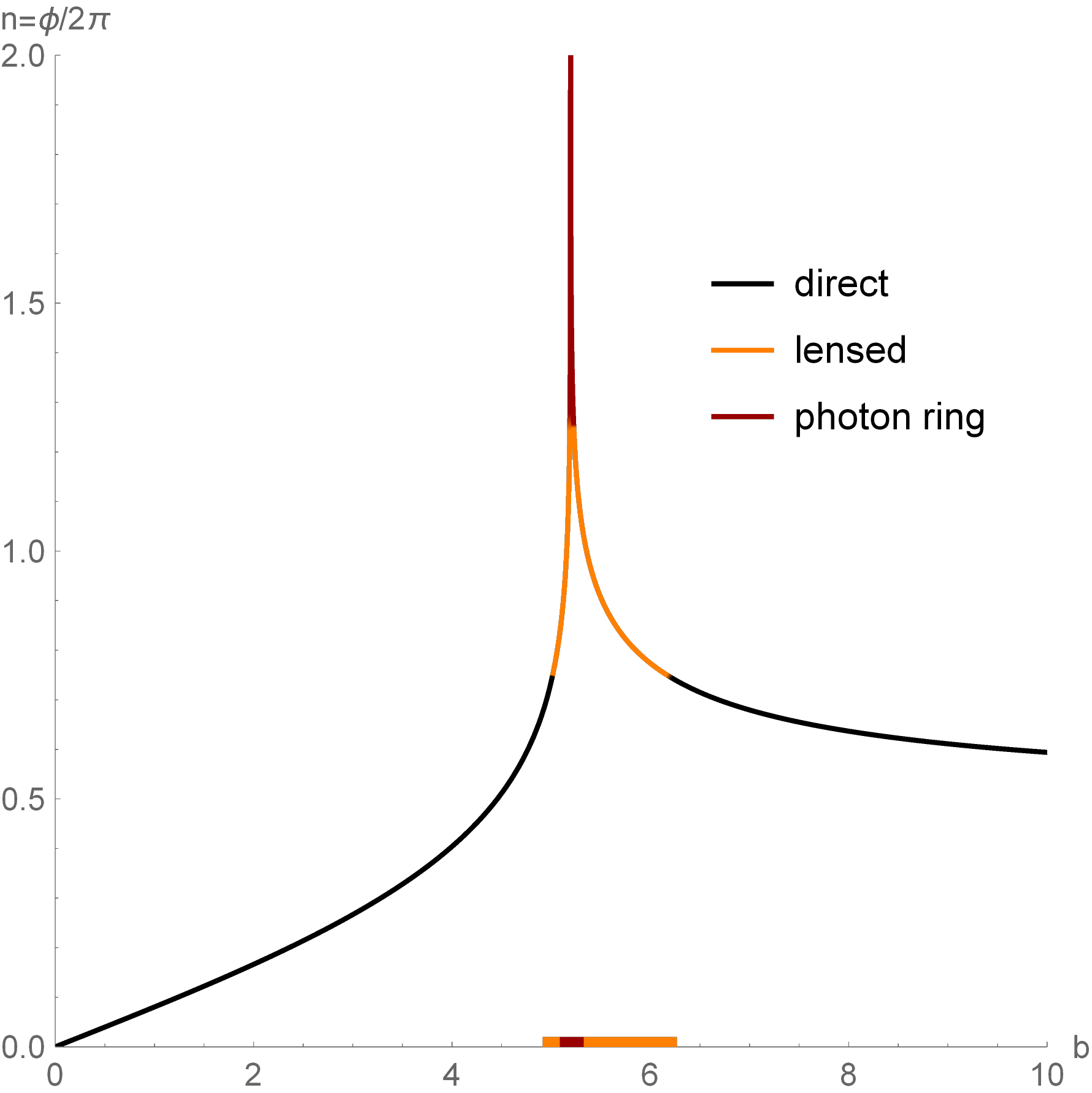}
    	\end{subfigure}
    	\caption{The dependence of the orbit number $n = \phi/(2\pi)$ on the impact parameter $b$ is shown for photons in Schwarzschild spacetime. Photon ring, lensed, and direct emissions are indicated by red, orange, and black curves, respectively.}
    	\label{ren}
    \end{figure}

\begin{figure*}[htbp]
    	\centering
    	\begin{subfigure}{0.38\textwidth}
    		\includegraphics[width=\linewidth]{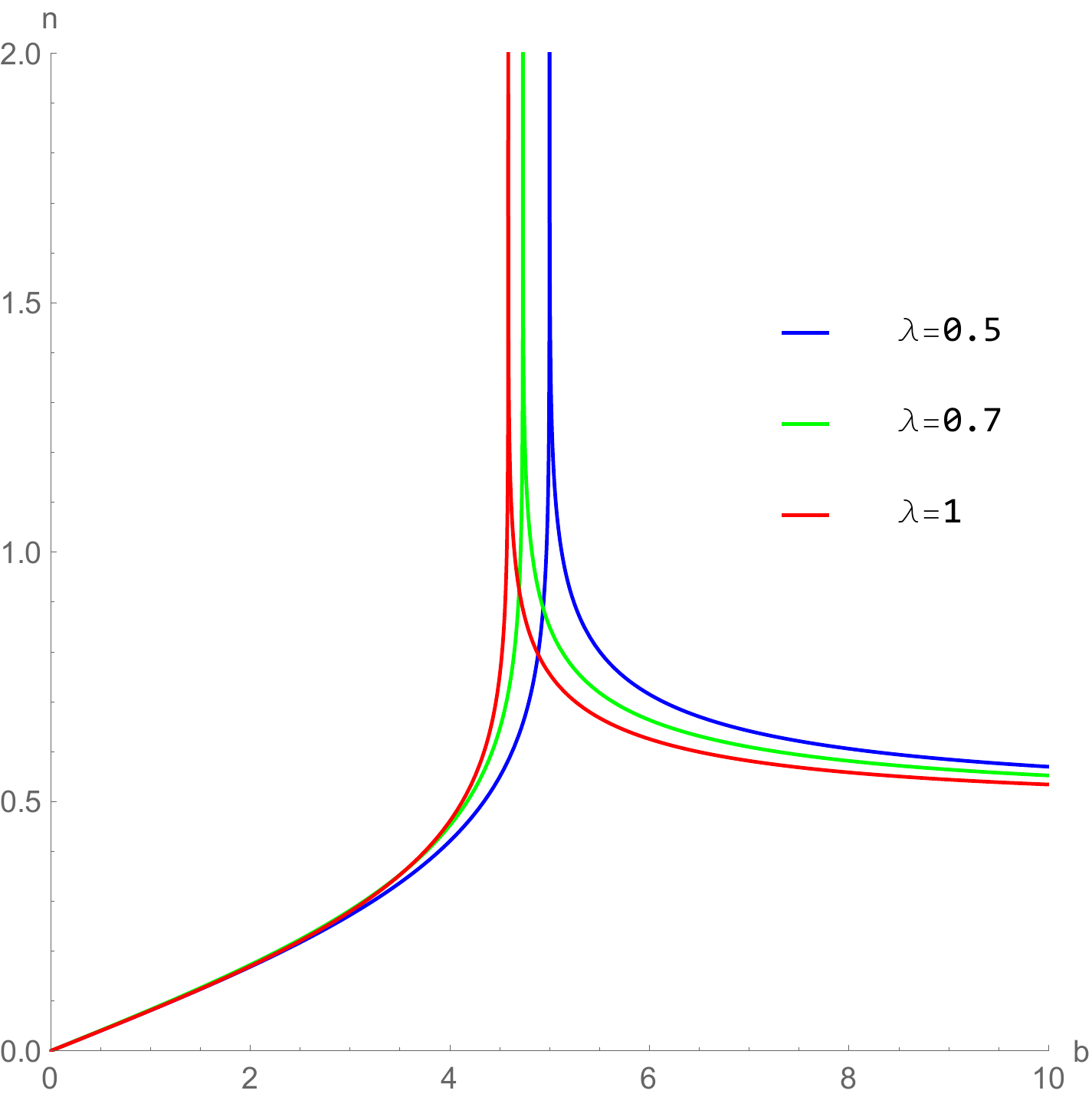}
    	\end{subfigure}%
    	\hspace{0.13\textwidth}%
    	\begin{subfigure}{0.38\textwidth}
    		\includegraphics[width=\linewidth]{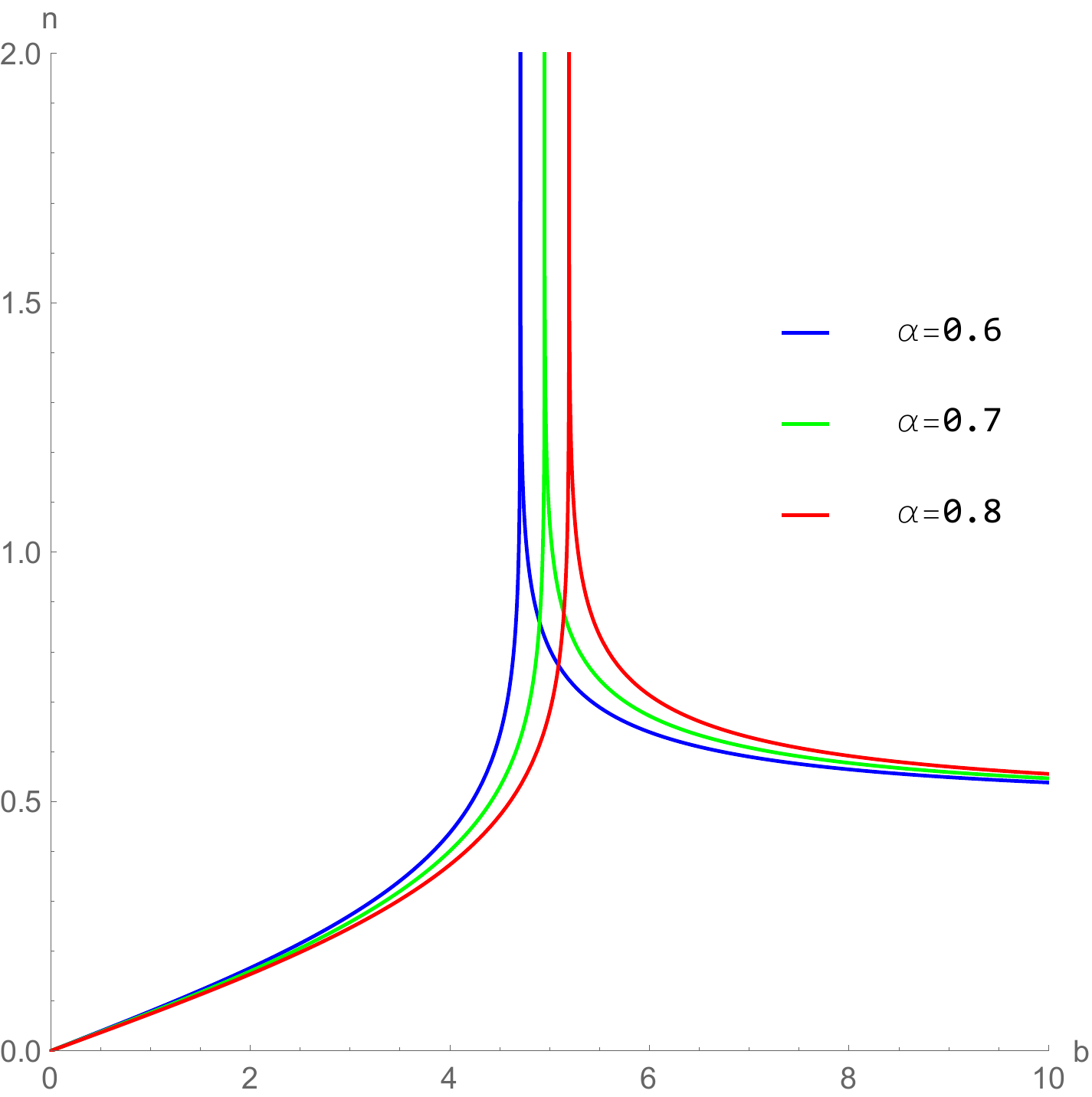}
    	\end{subfigure}
    	
    	\caption{The dependence of the orbit number $n = \phi/(2\pi)$ on the impact parameter $b$ for a black hole in STVG with PFDM. Left: $\alpha = 0.55$ is kept fixed and $\lambda$ varies as $0.5$, $0.7$, and $1$. Right: $\lambda = 1$ is held constant while $\alpha$ takes the values $0.6$, $0.7$, and $0.8$.}
    	\label{nb}
    \end{figure*}

\begin{figure*}[htbp]
    	\centering
    	\subfloat[$\alpha = 0.55, \lambda = 0.5$]{\includegraphics[width=0.28\textwidth]{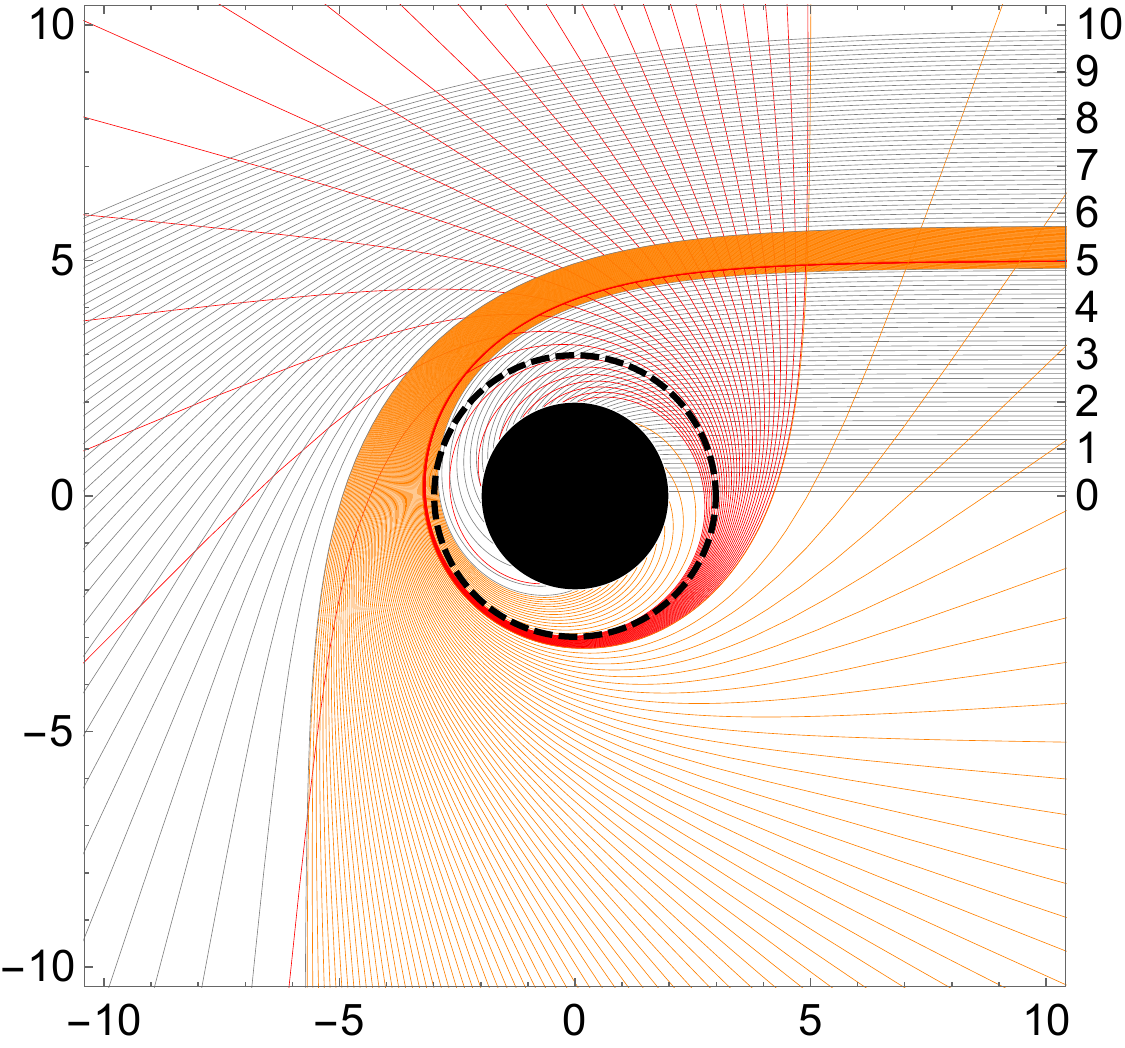}}\hfill
    	\subfloat[$\alpha = 0.55, \lambda=0.7$]{\includegraphics[width=0.28\textwidth]{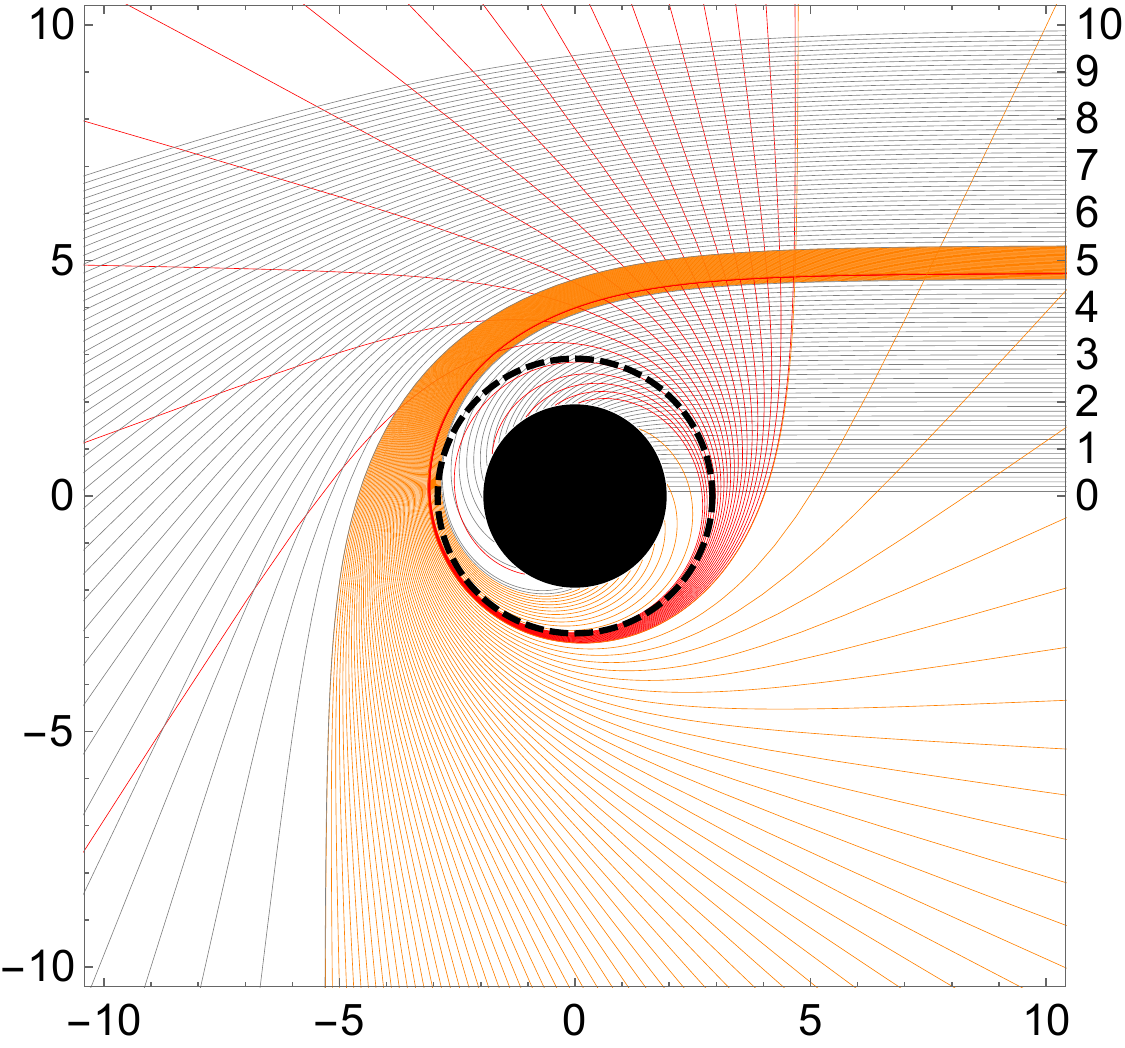}}\hfill
    	\subfloat[$\alpha = 0.55, \lambda=1$]{\includegraphics[width=0.28\textwidth]{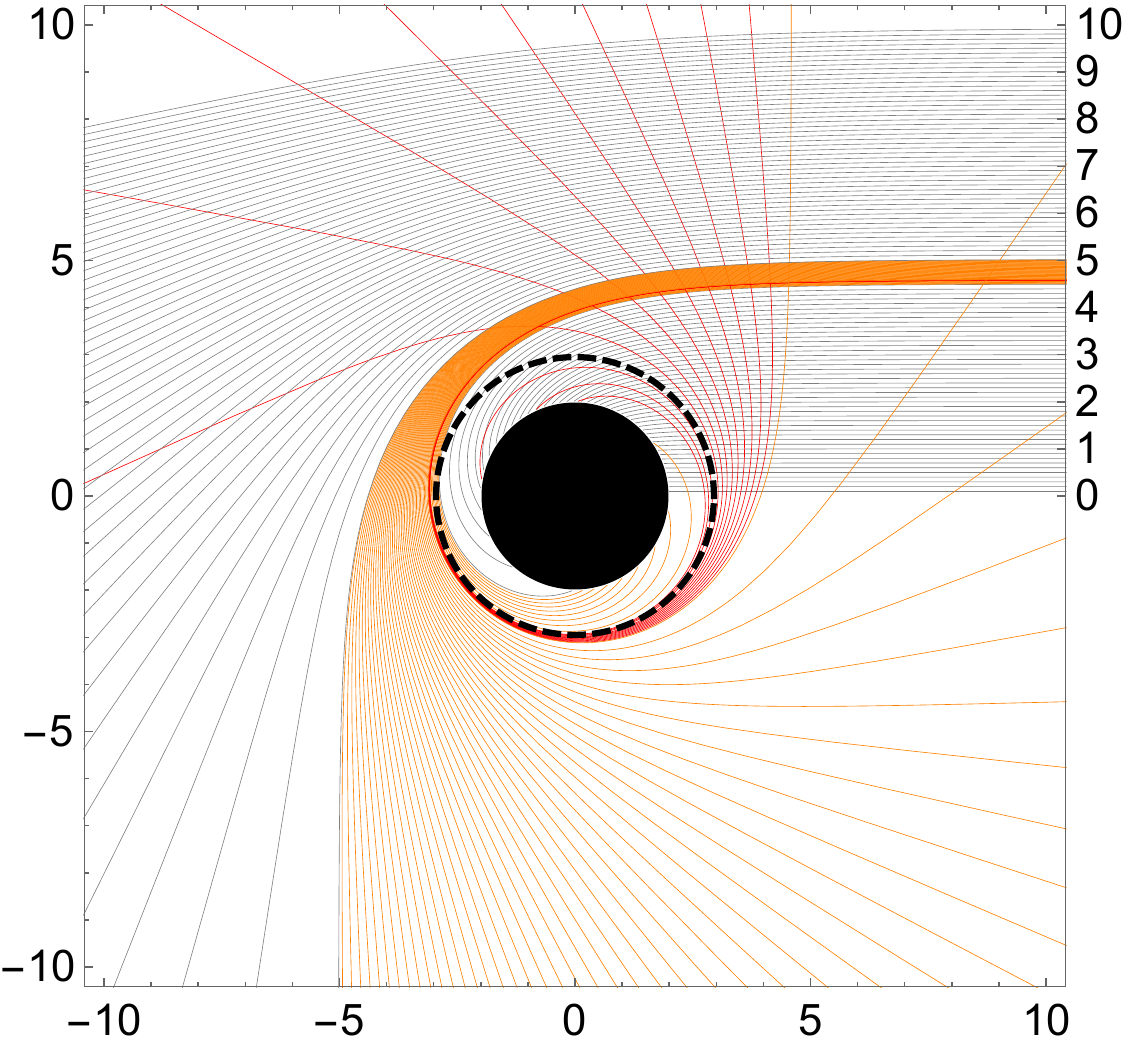}}\\
    	\subfloat[$\lambda = 1, \alpha=0.6$]{\includegraphics[width=0.28\textwidth]{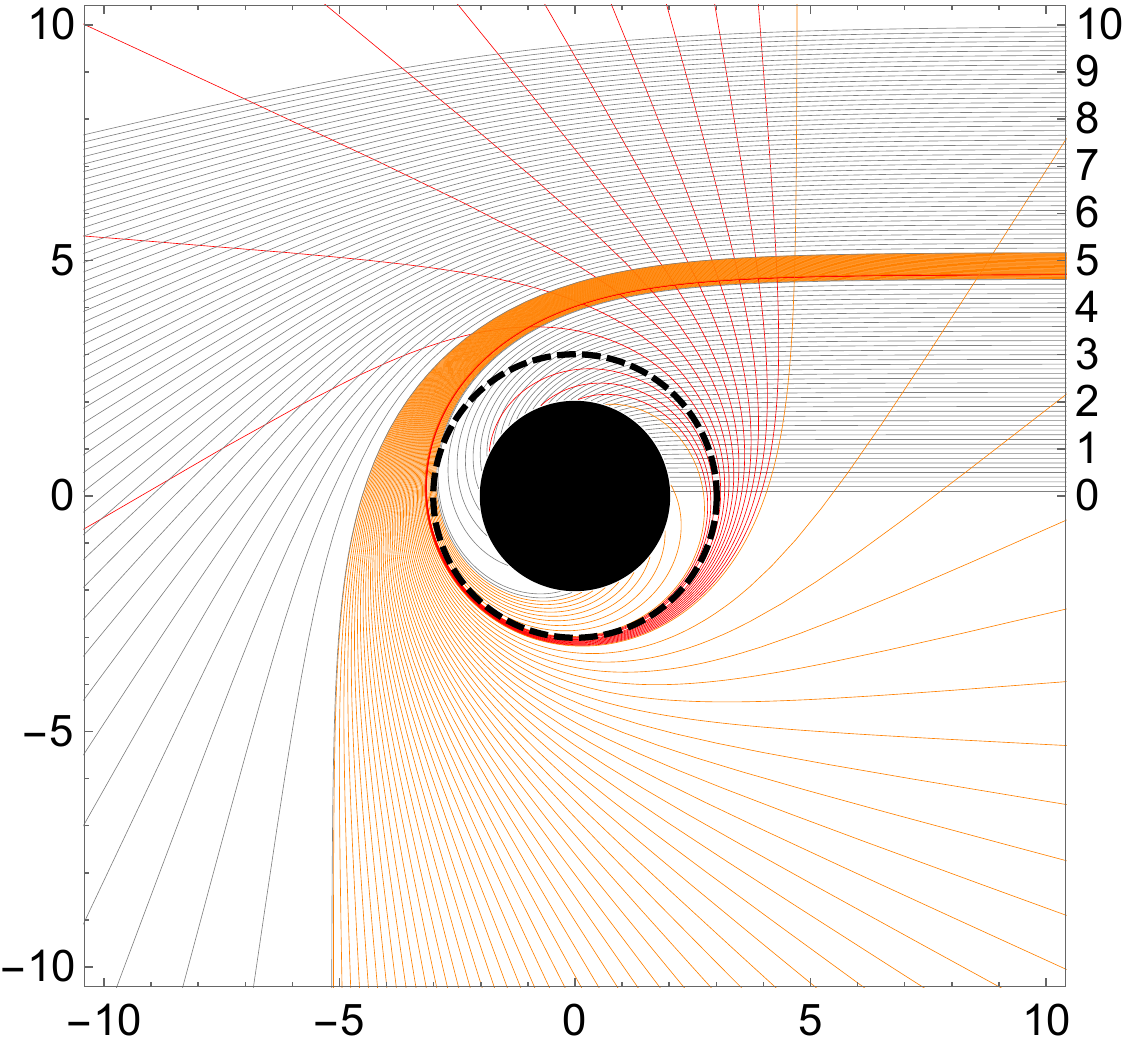}}\hfill
    	\subfloat[$\lambda = 1, \alpha=0.7$]{\includegraphics[width=0.28\textwidth]{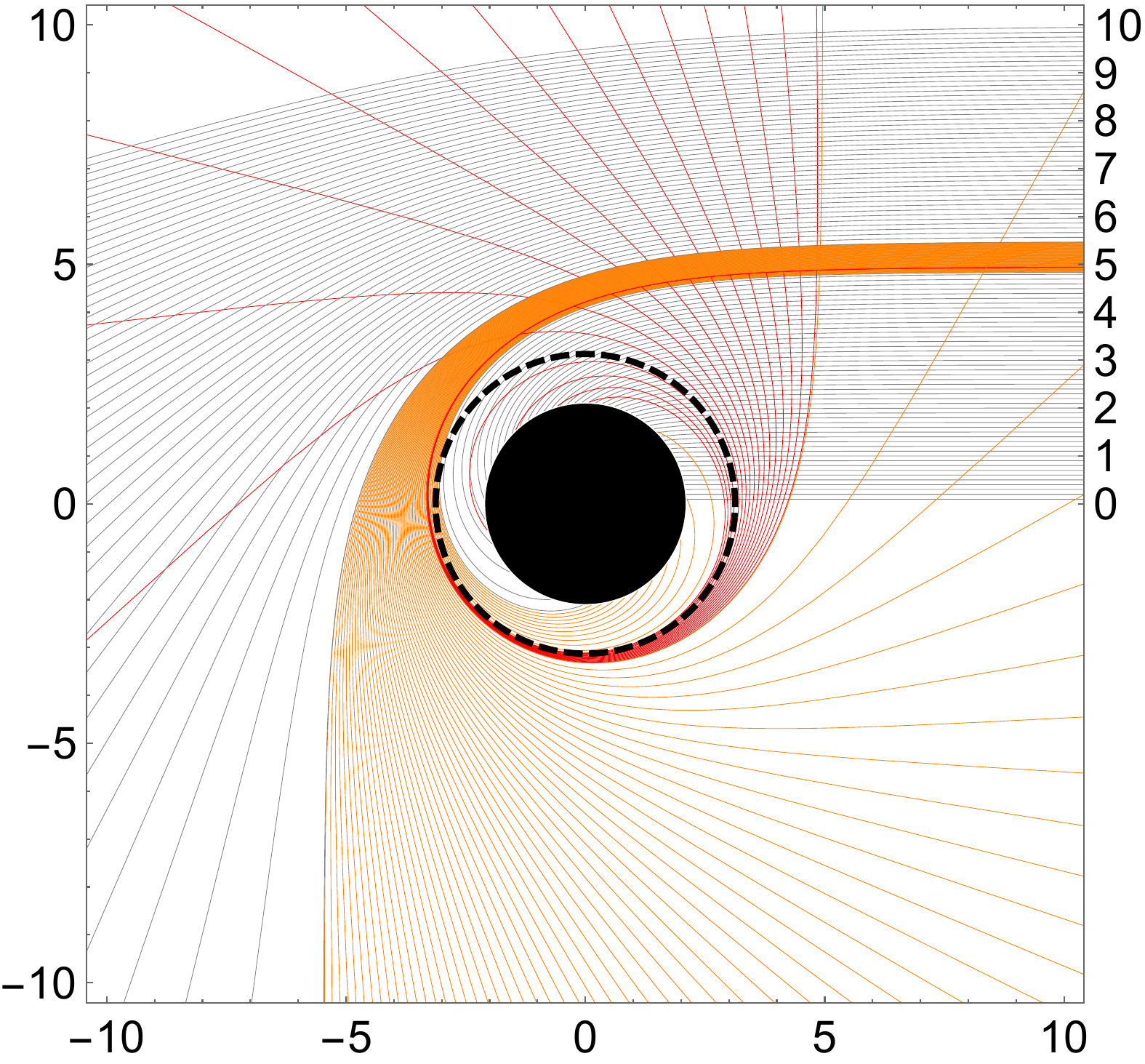}}\hfill
    	\subfloat[$\lambda = 1, \alpha=0.8$]{\includegraphics[width=0.28\textwidth]{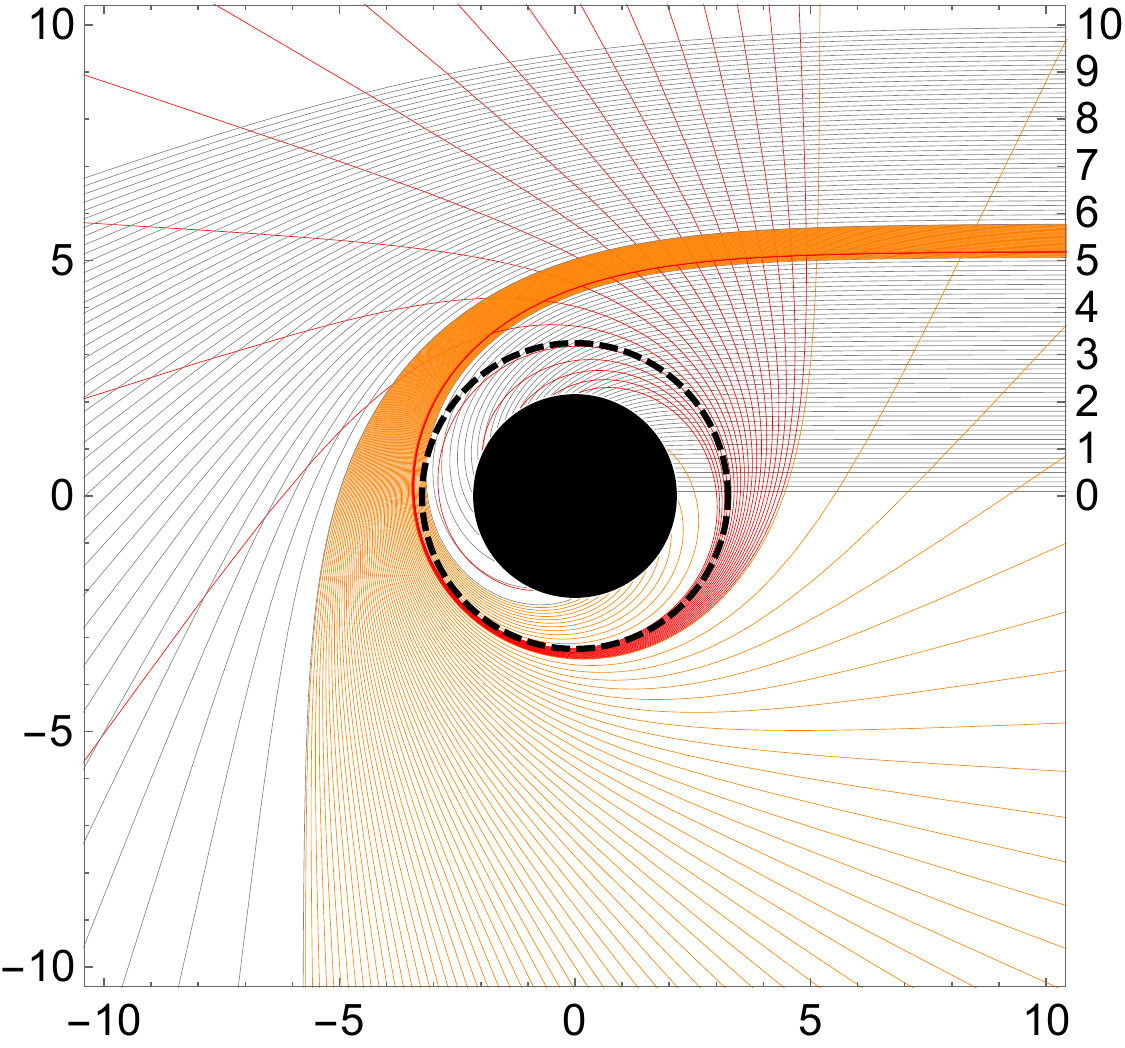}}
    	\caption{The photon trajectories around a black hole in STVG with PFDM are shown. Top: $\alpha = 0.55$ is kept constant while $\lambda$ varies: $\lambda = 0.5$, $0.7$, $1$. Bottom: $\lambda = 1$ is fixed, and $\alpha$ varies: $\alpha = 0.6$, $0.7$, $0.8$.}
    	\label{wanqu}
    \end{figure*}

    Photon trajectories originating from infinity are categorized according to the orbit number $n = \phi/(2\pi)$, as proposed in Ref.~\cite{Gralla:2019xty}. Trajectories with $n < 3/4$ are classified as direct emission, those with $3/4 < n < 5/4$ as lensed emission, and those with $n > 5/4$ as photon ring emission, based on the number of times the photon crosses the equatorial plane—namely, at most once, exactly twice, and three or more times, respectively.
    The dependence of the orbit number $n = \phi/(2\pi)$ on the impact parameter $b$ for photons in Schwarzschild geometry is shown in Fig.~\ref{ren}. Photon ring, lensed, and direct emission are indicated by the red, orange, and black curves, respectively.
    The dependence of the orbit number $n$ on the impact parameter $b$ is illustrated in Fig.~\ref{nb} for different choices of $\alpha$ and $\lambda$. In the right panel, with $\lambda = 1$ fixed, the curves shift to higher $b$ as $\alpha$ increases. In the left panel, for fixed $\alpha = 0.55$, the curves move toward lower $b$ with increasing $\lambda$.
    As illustrated in Fig.~\ref{wanqu}, the spatial distribution of lensed (orange) and photon ring (red) features in the black hole image is sensitive to both $\alpha$ and $\lambda$. Increasing the dark matter density (larger $\lambda$) at fixed $\alpha$ pulls these structures closer to the shadow edge and compresses them, whereas strengthening the modified gravity coupling (larger $\alpha$) at fixed $\lambda$ pushes them outward and broadens their width—effects that could, in principle, be disentangled through high-resolution imaging.

    \section{Quasinormal Modes and Their Connection to the Shadow}
    \label{qnms}
	Beyond their role in gravitational wave physics and stability analysis, QNMs have recently been linked to the optical appearance of black holes—particularly the shadow. Motivated by this correspondence, we compute the QNMs of a black hole in STVG with PFDM and explicitly demonstrate how the real part of the fundamental mode correlates with the shadow radius.
	
\begin{figure*}[htbp]
		\centering
		\begin{subfigure}{0.32\textwidth}
			\includegraphics[width=\textwidth, height=5.5in, keepaspectratio]{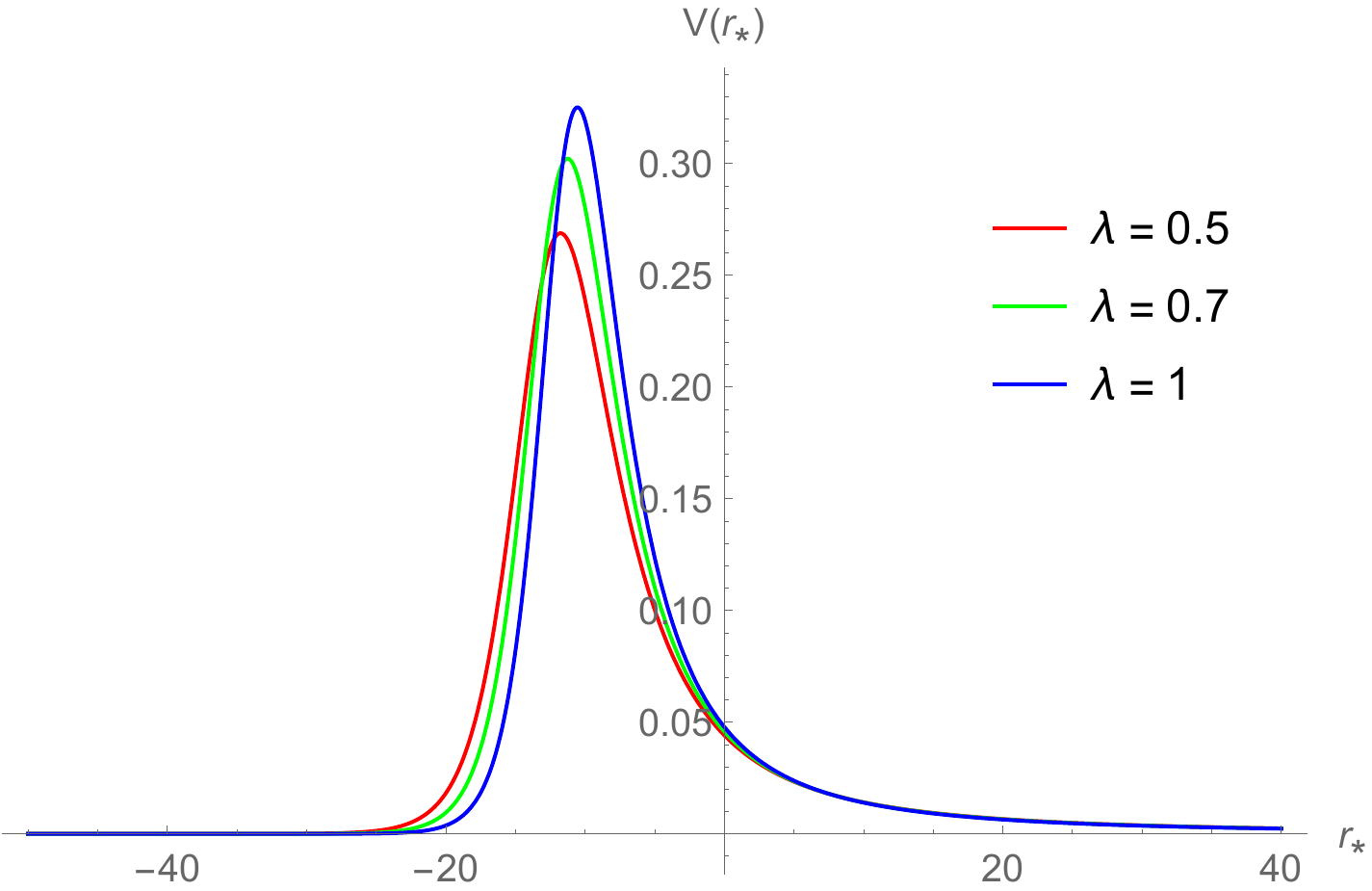}
		\end{subfigure}
		\hfill
		\begin{subfigure}{0.32\textwidth}
			\includegraphics[width=\textwidth, height=5.5in, keepaspectratio]{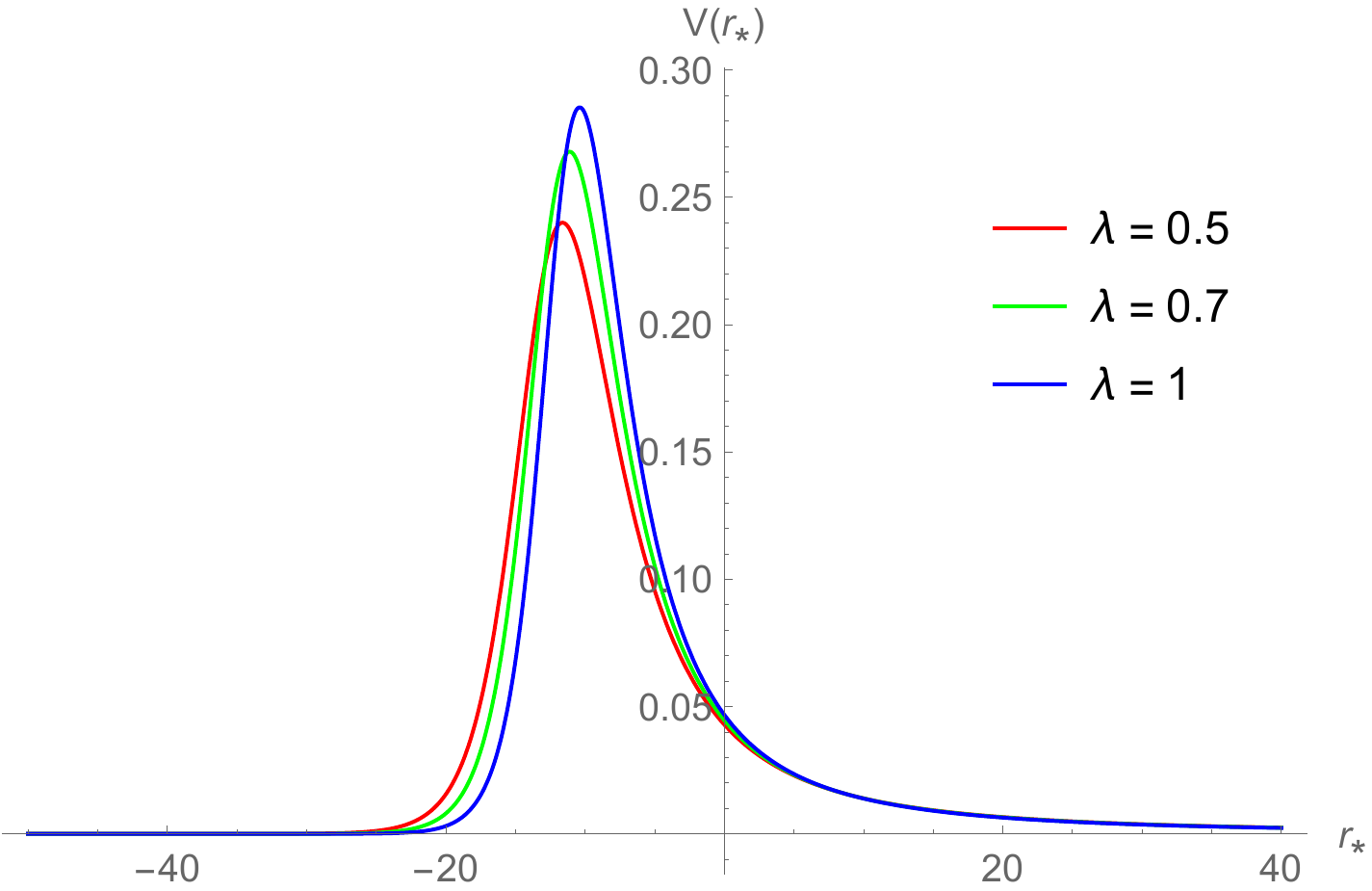}
		\end{subfigure}
		\hfill
		\begin{subfigure}{0.32\textwidth}
			\includegraphics[width=\textwidth, height=5.5in, keepaspectratio]{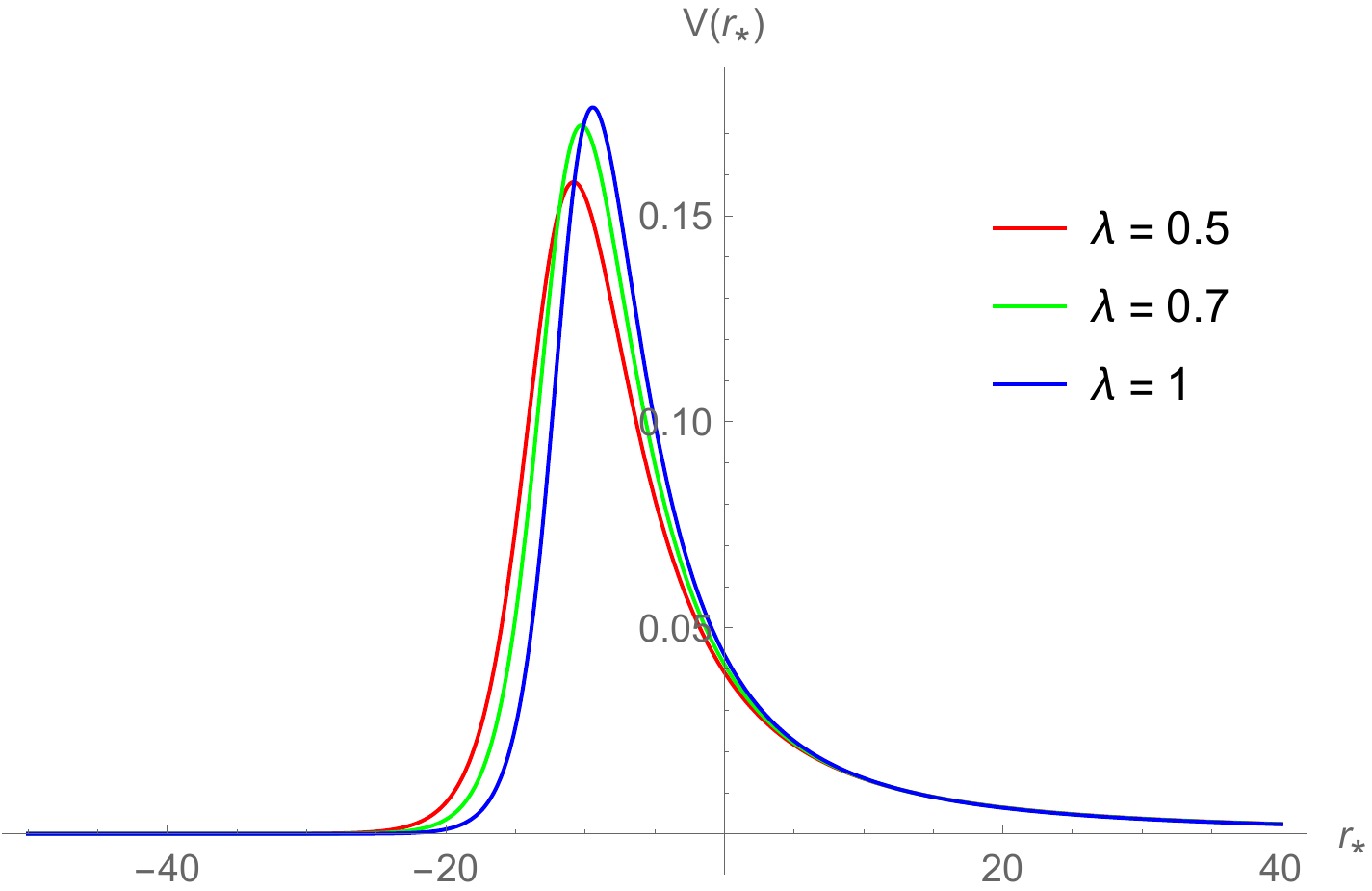}
		\end{subfigure}
		
		\vspace{0.3cm} 
		
		\begin{subfigure}{0.32\textwidth}
			\includegraphics[width=\textwidth, height=5.5in, keepaspectratio]{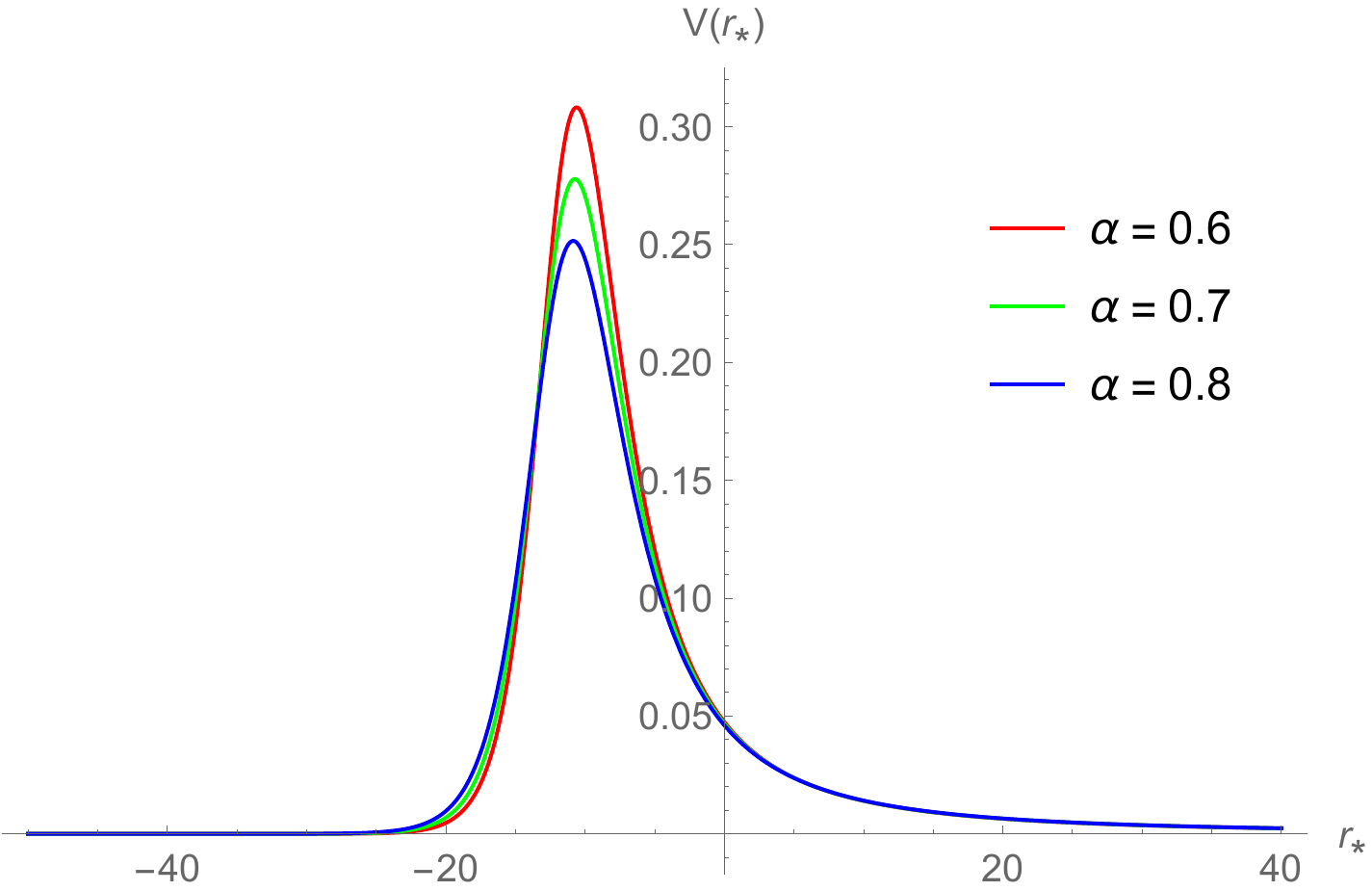}
		\end{subfigure}
		\hfill
		\begin{subfigure}{0.32\textwidth}
			\includegraphics[width=\textwidth, height=5.5in, keepaspectratio]{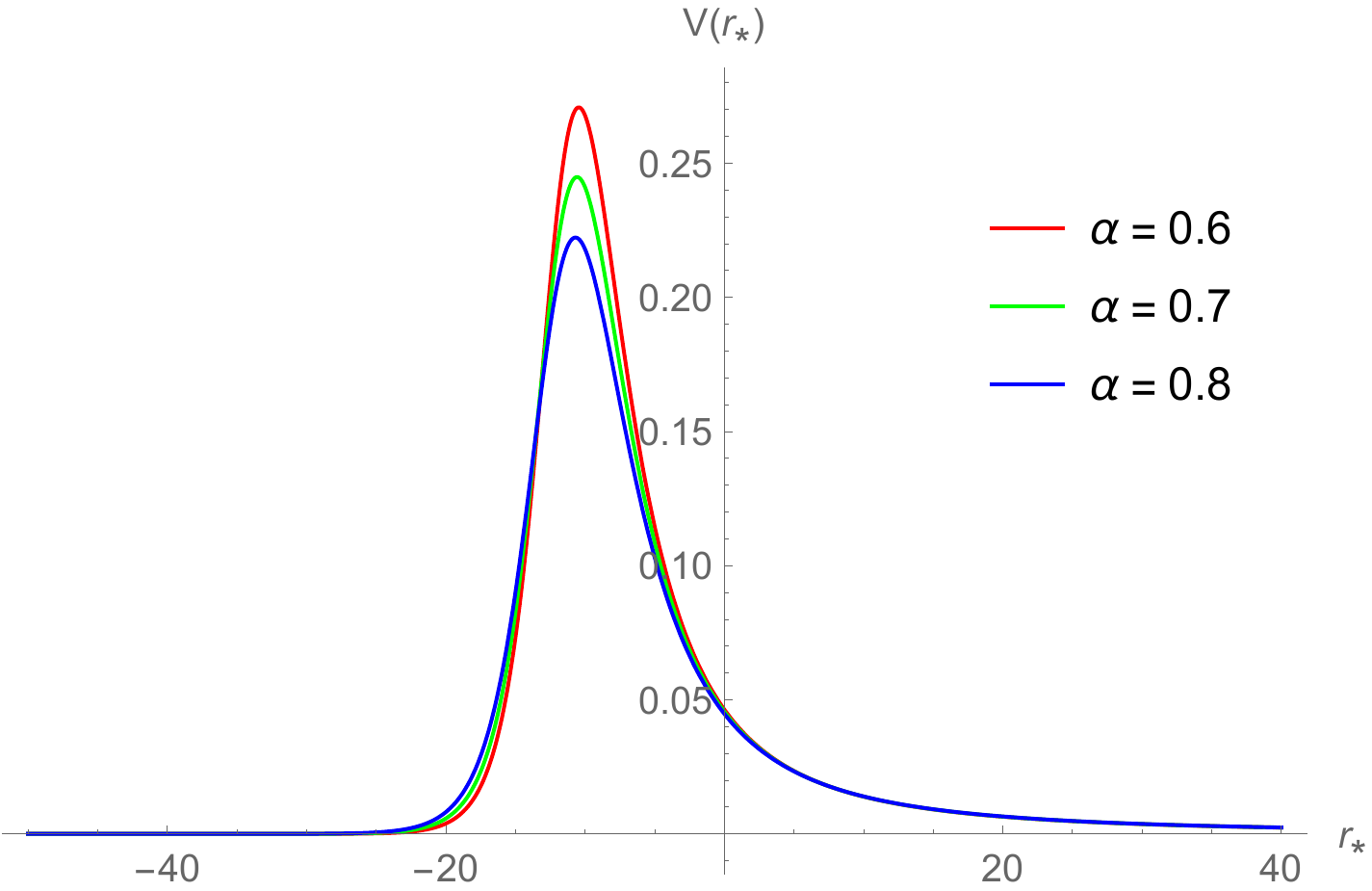}
		\end{subfigure}
		\hfill
		\begin{subfigure}{0.32\textwidth}
			\includegraphics[width=\textwidth, height=5.5in, keepaspectratio]{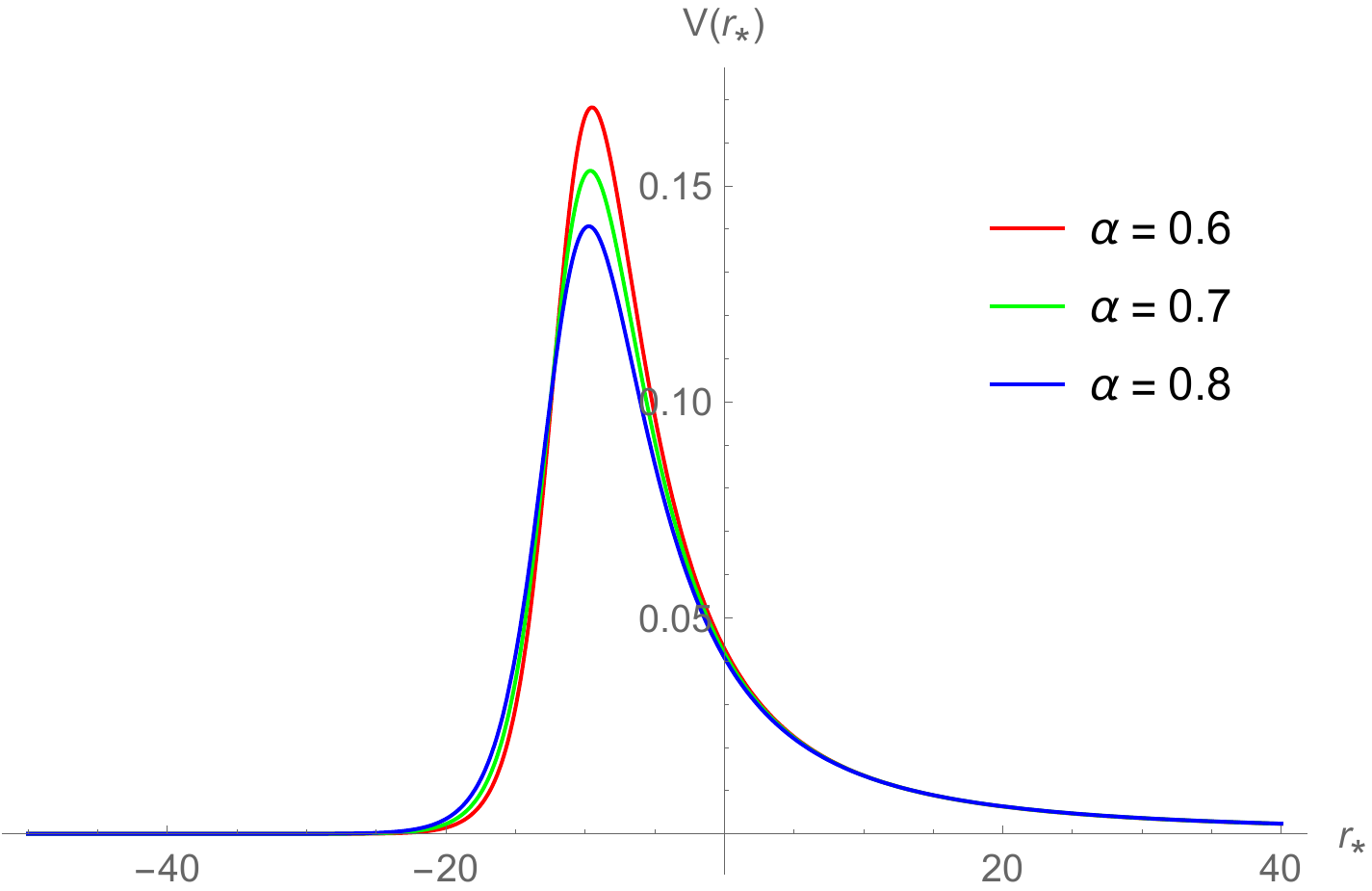}
		\end{subfigure}	
		\caption{The effective potential $V(r_*)$ is shown for $l=2$ perturbations: scalar ($s=0$, left), electromagnetic ($s=1$, middle), and axial gravitational ($s=2$, right). The top row presents results for $\alpha = 0.55$, while the bottom row corresponds to $\lambda = 1$.}
		\label{Vrstar}
	\end{figure*}

	By virtue of the spherical symmetry of the background geometry, the perturbation field $\Phi$ is expressed in terms of spherical harmonics as $\Phi(t,r,\theta,\phi) = Y(\theta,\phi)\Psi(t,r)/r$. The radial-temporal function $\Psi(t,r)$ obeys a Schr\"odinger-type equation,
	\begin{equation}
		\frac{\partial^2 \Psi}{\partial t^2} + \left( -\frac{d^2}{dr_*^2} + V(r) \right) \Psi = 0,\label{scheq}
	\end{equation}
	where the tortoise coordinate $r_*$ is related to $r$ by $dr_* = dr/f(r)$, and the potential $V(r)$ is given by
	\begin{equation}
		V(r) = f(r) \left[ \frac{l(l+1)}{r^2} + \frac{1 - s^2}{r} \frac{df}{dr} \right].\label{vqnm}
	\end{equation}
	The parameter $l$ represents the multipole quantum number, whereas the spin weight $s$ of the perturbation field assumes the values $s = 0$, $1$, and $2$ for scalar, electromagnetic, and gravitational cases, respectively. The effective potential shown in Fig. \ref{Vrstar} is found to depend sensitively on $\alpha$ and $\lambda$: the peak value increases with $\lambda$ and decreases with $\alpha$. Additionally, for given values of these parameters, the peak is observed to be highest in the case of scalar perturbations ($s=0$), followed by electromagnetic ($s=1$), and then gravitational ($s=2$) perturbations. A diminished potential barrier suggests less effective confinement of linear perturbations, thereby weakening the system's resilience to external perturbations.
	By adopting the time-harmonic ansatz $\Psi(t,r) = e^{-i\omega t}\psi(r)$, the time-dependent wave equation is transformed into the stationary form
	\begin{equation}
		\frac{d^2\psi}{dr_*^2} + \left[ \omega^2 - V(r) \right] \psi = 0,\label{dpsi}
	\end{equation}
	in which $\omega$ represents the complex frequency associated with the QNMs of the black hole spacetime.
	
	The complex frequencies $\omega$ are obtained via the WKB approximation, wherein the wave equation is interpreted as describing tunneling across a potential barrier in analogy with quantum mechanics. An approximate analytical solution is derived from the following condition:
	\begin{equation}
		\frac{i(\omega^2 - V_0)}{\sqrt{-2 V_0''}} - \sum_{i=2}^{6} \Lambda_i = n + \frac{1}{2}, \quad (n = 0,1,2,\ldots),\label{wkbeq}
	\end{equation}
	The quantity $V_0$ is defined as the maximum of the effective potential, while $\Lambda_i$ corresponds to the $i$-th order correction in the WKB expansion. The integer $n$ denotes the overtone number, with different values of $n$ associated with distinct quasinormal oscillation modes. As shown in Tables \ref{tabWKB} and \ref{tabWKB2}, the fundamental QNM spectrum exhibits a clear parametric dependence: increasing $\lambda$ enhances both the oscillation frequency ($\operatorname{Re}\,\omega$) and damping rate ($|\operatorname{Im}\,\omega|$), whereas larger $\alpha$ suppresses both. Notably, across all parameter choices, scalar perturbations ($s=0$) consistently produce the highest-frequency and most strongly damped modes, followed by electromagnetic ($s=1$) and then axial gravitational ($s=2$) perturbations—a hierarchy directly inherited from the structure of the effective potential.

	\begin{table*}[!htbp]  
		\setlength{\abovecaptionskip}{0.2cm}
		\setlength{\belowcaptionskip}{0.2cm}
		\centering
		\caption{The fundamental QNM frequencies $\omega$ are tabulated for $l=2$ perturbations of spin weight $s=0$, $1$, and $2$ in a black hole spacetime described by STVG with PFDM. The sixth-order WKB approximation is employed with $\alpha = 0.55$, while $\lambda$ is varied.}
		\label{tabWKB}
		\resizebox{\textwidth}{!}{
			\begin{tabular}{lcccccc}
				\hline 
				& 1st-order & 2nd-order & 3rd-order & 4th-order & 5th-order & 6th-order \\
				\hline
				$s = 0$ & \multicolumn{6}{c}{$\alpha = 0.55$} \\
				\hline
				$\lambda=0.5$ & $  0.528627 -0.102979i$ & $  0.503996 -0.108012i$ & $  0.503084 -0.103672i$ & $  0.503659 -0.103554i$ & $  0.503678 -0.103644i$ & $  0.503644 -0.103651i$ \\
				$\lambda=0.7$ & $  0.561385 -0.113857i$ & $  0.532911 -0.11994i$ & $  0.53175 -0.114671i$ & $  0.532479 -0.114514i$ & $  0.532504 -0.114632i$ & $  0.532458 -0.114642i$ \\
				$\lambda=1$ & $  0.583798 -0.125118i$ & $  0.550534 -0.132678i$ & $  0.548983 -0.126089i$ & $  0.54995 -0.125868i$ & $  0.549987 -0.126029i$ & $  0.549918 -0.126045i$ \\
				\hline
				$s = 1$ & \multicolumn{6}{c}{$\alpha = 0.55$} \\
				\hline
				$\lambda=0.5$ & $ 0.500191 -0.100849i$ & $ 0.475084 -0.106179i$ & $ 0.474073 -0.101559i$ & $ 0.474701 -0.101425i$ & $ 0.474722 -0.101522i$ & $ 0.474684 -0.10153i$ \\
				$\lambda=0.7$ & $ 0.529427 -0.111235i$ & $ 0.500435 -0.11768i$ & $ 0.499146 -0.112073i$ & $ 0.499937 -0.111896i$ & $ 0.499964 -0.112017i$ & $ 0.499914 -0.112028i$ \\
				$\lambda=1$ & $ 0.547858 -0.121791i$ & $ 0.514053 -0.1298i$ & $ 0.51233 -0.122798i$ & $ 0.513364 -0.122551i$ & $  0.513401 -0.122704i$ & $ 0.513336 -0.12272i$ \\
				\hline
				$s = 2$ & \multicolumn{6}{c}{$\alpha = 0.55$} \\
				\hline
				$\lambda=0.5$ & $ 0.408476 -0.0931762i$ & $ 0.381676 -0.0997189i$ & $ 0.380229 -0.0940286i$ & $ 0.380932 -0.093855i$ & $ 0.380912 -0.0937729i$ & $ 0.381072 -0.0937335i$ \\
				$\lambda=0.7$ & $ 0.42708 -0.102289i$ & $ 0.395985 -0.110321i$ & $ 0.394123 -0.10344i$ & $ 0.394756 -0.103274i$ & $ 0.394656 -0.102891i$ & $ 0.394917 -0.102823i$ \\
				$\lambda=1$ & $ 0.434409 -0.111447i$ & $ 0.397935 -0.121663i$ & $ 0.395484 -0.113391i$ & $ 0.395614 -0.113354i$ & $ 0.39538 -0.112532i$ & $ 0.395041 -0.112628i$ \\
				\hline 
			\end{tabular}%
		}
	\end{table*}
	
	\begin{table*}[!htbp]  
		\setlength{\abovecaptionskip}{0.2cm}
		\setlength{\belowcaptionskip}{0.2cm}
		\centering
		\caption{The fundamental QNM frequencies $\omega$ are tabulated for $l=2$ perturbations of spin weight $s=0$, $1$, and $2$ in a black hole spacetime described by STVG with PFDM. The sixth-order WKB approximation is employed with $\lambda = 1$, while $\alpha$ is varied.}
		\label{tabWKB2}
		\resizebox{\textwidth}{!}{
			\begin{tabular}{lcccccc}
				\hline 
				& 1st-order & 2nd-order & 3rd-order & 4th-order & 5th-order & 6th-order \\
				\hline
				$s = 0$ & \multicolumn{6}{c}{$\lambda = 1$} \\
				\hline
				$\alpha=0.6$ & $  0.568087 -0.120663i$ & $  0.536329 -0.127808i$ & $  0.53488 -0.121584i$ & $  0.535788 -0.121378i$ & $  0.535822 -0.121531i$ & $  0.535758 -0.121546i$ \\
				$\alpha=0.7$ & $  0.538909 -0.112448i$ & $  0.509893 -0.118847i$ & $  0.508623 -0.113276i$ & $  0.509429 -0.113097i$ & $  0.509459 -0.113235i$ & $  0.509403 -0.113248i$ \\
				$\alpha=0.8$ & $  0.512385 -0.105052i$ & $  0.485797 -0.110802i$ & $  0.484679 -0.105793i$ & $  0.4854 -0.105636i$ & $  0.485428 -0.105763i$ & $  0.485377 -0.105774i$ \\
				\hline
				$s = 1$ & \multicolumn{6}{c}{$\lambda = 1$} \\
				\hline
				$\alpha=0.6$ & $ 0.5335 -0.117522i$ & $ 0.501211 -0.125093i$ & $ 0.4996 -0.118475i$ & $ 0.500577 -0.118243i$ & $ 0.500613 -0.118393i$ & $ 0.50055 -0.118408i$ \\
				$\alpha=0.7$ & $ 0.506799 -0.109641i$ & $ 0.477276 -0.116423i$ & $ 0.475863 -0.11049i$ & $ 0.476738 -0.110288i$ & $ 0.476771 -0.110431i$ & $ 0.476709 -0.110445i$ \\
				$\alpha=0.8$ & $ 0.482485 -0.102532i$ & $ 0.455415 -0.108627i$ & $ 0.454171 -0.103287i$ & $ 0.454961 -0.103108i$ & $ 0.454992 -0.103245i$ & $ 0.454933 -0.103259i$ \\
				\hline
				$s = 2$ & \multicolumn{6}{c}{$\lambda = 1$} \\
				\hline
				$\alpha=0.6$ & $ 0.423937 -0.107546i$ & $ 0.389132 -0.117165i$ & $ 0.386821 -0.109247i$ & $ 0.387093 -0.10917i$ & $ 0.38688 -0.108412i$ & $ 0.386773 -0.108442i$ \\
				$\alpha=0.7$ & $ 0.404413 -0.100342i$ & $ 0.372672 -0.108888i$ & $ 0.37062 -0.101643i$ & $ 0.371134 -0.101502i$ & $ 0.370977 -0.100925i$ & $ 0.371193 -0.100866i$ \\
				$\alpha=0.8$ & $ 0.386577 -0.0938418i$ & $ 0.357574 -0.101454i$ & $ 0.355753 -0.0948381i$ & $ 0.356455 -0.0946513i$ & $ 0.356362 -0.0942985i$ & $ 0.356714 -0.0942053i$ \\
				\hline 
			\end{tabular}%
		}
	\end{table*}
	
	The Pad\'e approximant $P_N^M(x)$ of a formal power series $\sum_{k=0}^\infty a_k x^k$ is defined as the unique rational function whose numerator and denominator have degrees at most $M$ and $N$, respectively, and which matches the series up to order $M+N$:
	\begin{equation}
		P_N^M(x) - \sum_{k=0}^{M+N} a_k x^k = \mathcal{O}(x^{M+N+1}).
		\label{pade}
	\end{equation}
	As defined by Eq. (\ref{pade}), the Pad\'e approximant $P_N^M(x)$ provides a rational approximation to a power series that is accurate through order $M+N$. Applying this technique to the QNM problem yields frequencies (see Tables \ref{tabPADE1}-\ref{tabPADE6}, $k=1$–$13$) that exhibit remarkable consistency with the sixth-order WKB predictions, thereby reinforcing confidence in the computed spectrum.
			
			\begin{table*}[!htbp]  
				\setlength{\abovecaptionskip}{0.2cm}
				\setlength{\belowcaptionskip}{0.2cm}
				\centering
				\caption{The QNM frequencies $\omega$ are presented for scalar perturbations ($s=0$, $l=2$) in a black hole spacetime described by STVG with PFDM. Results are obtained via Pad\'e approximants of order $k = 1$ to $13$, with $\alpha$ fixed at $0.55$ and $\lambda$ varied.}
				\label{tabPADE1}
				\resizebox{\textwidth}{!}{
					\begin{tabular}{lccc}
						\hline 
						$\alpha = 0.55$ & $\lambda = 0.5$ & $\lambda = 0.7$ & $\lambda = 1$  \\
						\hline
						$k=1$ & $   0.509299 -0.0992139i$ & $   0.539205 -0.109358i$ & $   0.558161 -0.119623i$  \\
						$k=2$ & $   0.503631 -0.103122i$ & $   0.53246 -0.113978i$ & $   0.549956 -0.125179i$  \\
						$k=3$ & $   0.503558 -0.103562i$ & $   0.532355 -0.114523i$ & $   0.549796 -0.125878i$  \\						
						$k=4$ & $   0.50366 -0.103626i$ & $   0.532481 -0.114608i$ & $   0.549954 -0.125996i$  \\
						$k=5$ & $   0.503656 -0.103642i$ & $   0.532475 -0.114628i$ & $   0.549945 -0.126023i$  \\
						$k=6$ & $   0.503647 -0.103636i$ & $   0.532462 -0.11462i$ & $   0.549928 -0.12601i$  \\
						$k=7$ & $   0.503647 -0.103637i$ & $   0.532467 -0.114621i$ & $   0.549924 -0.126013i$  \\
						$k=8$ & $   0.503647 -0.103636i$ & $   0.532463 -0.11462i$ & $   0.549927 -0.12601i$  \\
						$k=9$ & $   0.503647 -0.103636i$ & $   0.532466 -0.11462i$ & $   0.549921 -0.126026i$  \\
						$k=10$ & $   0.503649 -0.103636i$ & $   0.532465 -0.11462i$ & $   0.54993 -0.126011i$  \\
						$k=11$ & $   0.503655 -0.103647i$ & $   0.532472 -0.114621i$ & $   0.549939 -0.126012i$  \\
						$k=12$ & $   0.503649 -0.103636i$ & $   0.532464 -0.11462i$ & $   0.54993 -0.126011i$  \\
						$k=13$ & $   0.503646 -0.103638i$ & $   0.532472 -0.114621i$ & $   0.549944 -0.126033i$  \\
						
						\hline 
					\end{tabular}%
				}
			\end{table*}
			
			\begin{table*}[!htbp]  
				\setlength{\abovecaptionskip}{0.2cm}
				\setlength{\belowcaptionskip}{0.2cm}
				\centering
				\caption{The QNM frequencies $\omega$ are presented for electromagnetic perturbations ($s=1$, $l=2$) in a black hole spacetime described by STVG with PFDM. Results are obtained via Pad\'e approximants of order $k = 1$ to $13$, with $\alpha$ fixed at $0.55$ and $\lambda$ varied.}
				\label{tabPADE2}
				\resizebox{\textwidth}{!}{
					\begin{tabular}{lccc}
						\hline 
						$\alpha = 0.55$ & $\lambda = 0.5$ & $\lambda = 0.7$ & $\lambda = 1$  \\
						\hline
						$k=1$ & $    0.480652 -0.0969096i$ & $    0.507044 -0.106533i$ & $    0.522059 -0.116055i$  \\
						$k=2$ & $    0.474698 -0.10094i$ & $    0.499958 -0.111287i$ & $    0.513443 -0.121753i$  \\
						$k=3$ & $    0.474597 -0.101429i$ & $    0.499813 -0.111898i$ & $    0.513219 -0.122544i$  \\						
						$k=4$ & $    0.474704 -0.101502i$ & $    0.499941 -0.111992i$ & $    0.513369 -0.12267i$  \\
						$k=5$ & $    0.474698 -0.101519i$ & $    0.499934 -0.112013i$ & $    0.513363 -0.122698i$  \\
						$k=6$ & $    0.474688 -0.101512i$ & $    0.499922 -0.112003i$ & $    0.513348 -0.122684i$  \\
						$k=7$ & $    0.474695 -0.101514i$ & $    0.499929 -0.112007i$ & $    0.513353 -0.122686i$  \\
						$k=8$ & $    0.474688 -0.101511i$ & $    0.499921 -0.112004i$ & $    0.513347 -0.122684i$  \\
						$k=9$ & $    0.474695 -0.101515i$ & $    0.49993 -0.112007i$ & $    0.513358 -0.122691i$  \\
						$k=10$ & $    0.474689 -0.101509i$ & $    0.499919 -0.112006i$ & $    0.513346 -0.122691i$  \\
						$k=11$ & $    0.474699 -0.101514i$ & $    0.499933 -0.112006i$ & $    0.513364 -0.122691i$  \\
						$k=12$ & $    0.47469 -0.10151i$ & $    0.499919 -0.112006i$ & $    0.513345 -0.12269i$  \\
						$k=13$ & $    0.474699 -0.101515i$ & $    0.499934 -0.112007i$ & $    0.513366 -0.122695i$  \\
						
						\hline 
					\end{tabular}%
				}
			\end{table*}
			
			\begin{table*}[!htbp]  
				\setlength{\abovecaptionskip}{0.2cm}
				\setlength{\belowcaptionskip}{0.2cm}
				\centering
				\caption{The QNM frequencies $\omega$ are presented for axial gravitational perturbations ($s=2$, $l=2$) in a black hole spacetime described by STVG with PFDM. Results are obtained via Pad\'e approximants of order $k = 1$ to $13$, with $\alpha$ fixed at $0.55$ and $\lambda$ varied.}
				\label{tabPADE3}
				\resizebox{\textwidth}{!}{
					\begin{tabular}{lccc}
						\hline 
						$\alpha = 0.55$ & $\lambda = 0.5$ & $\lambda = 0.7$ & $\lambda = 1$  \\
						\hline
						$k=1$ & $     0.388274 -0.0885678i$ & $     0.403911 -0.0967393i$ & $     0.407583 -0.104565i$  \\
						$k=2$ & $     0.381208 -0.0930199i$ & $     0.3954 -0.101997i$ & $     0.397129 -0.11098i$  \\
						$k=3$ & $     0.380914 -0.0937773i$ & $     0.394897 -0.103049i$ & $     0.396082 -0.11264i$  \\						
						$k=4$ & $     0.380972 -0.0938262i$ & $     0.394867 -0.103023i$ & $     0.395755 -0.112936i$  \\
						$k=5$ & $     0.380943 -0.0937807i$ & $     0.394916 -0.103012i$ & $     0.395862 -0.11256i$  \\
						$k=6$ & $     0.381002 -0.0938297i$ & $     0.394917 -0.102926i$ & $     0.395088 -0.112699i$  \\
						$k=7$ & $     0.380998 -0.0938228i$ & $     0.394898 -0.10295i$ & $     0.396046 -0.112124i$  \\
						$k=8$ & $     0.381005 -0.093829i$ & $     0.394921 -0.102925i$ & $     0.395624 -0.112695i$  \\
						$k=9$ & $     0.381009 -0.0938376i$ & $     0.394917 -0.103009i$ & $     0.396042 -0.112124i$  \\
						$k=10$ & $     0.381011 -0.0938296i$ & $     0.394925 -0.102925i$ & $     0.395515 -0.112545i$  \\
						$k=11$ & $     0.380995 -0.0938259i$ & $     0.394834 -0.103i$ & $     0.396058 -0.112129 i$  \\
						$k=12$ & $     0.38101 -0.0938289i$ & $     0.394921 -0.102924i$ & $     0.395515 -0.112546i$  \\
						$k=13$ & $     0.380993 -0.0938252i$ & $     0.394781 -0.102853i$ & $     0.395807 -0.112123i$  \\
						
						\hline 
					\end{tabular}%
				}
			\end{table*}

			\begin{table*}[!htbp]  
				\setlength{\abovecaptionskip}{0.2cm}
				\setlength{\belowcaptionskip}{0.2cm}
				\centering
				\caption{The QNM frequencies $\omega$ are presented for scalar perturbations ($s=0$, $l=2$) in a black hole spacetime described by STVG with PFDM. Results are obtained via Pad\'e approximants of order $k = 1$ to $13$, with $\lambda$ fixed at $1$ and $\alpha$ varied.}
				\label{tabPADE4}
				\resizebox{\textwidth}{!}{
					\begin{tabular}{lccc}
						\hline 
						$\lambda = 1$ & $\alpha = 0.6$ & $\alpha = 0.7$ & $\alpha = 0.8$  \\
						\hline
						$k=1$ & $    0.543565 -0.115454i$ & $    0.516425 -0.107757i$ & $    0.491715 -0.100815i$  \\
						$k=2$ & $    0.535785 -0.120736i$ & $    0.50941 -0.112539i$ & $    0.485368 -0.105154i$  \\
						$k=3$ & $    0.535641 -0.12139i$ & $    0.509295 -0.113112i$ & $    0.485277 -0.105654i$  \\						
						$k=4$ & $    0.535791 -0.121499i$ & $    0.509431 -0.113206i$ & $    0.485401 -0.105737i$  \\
						$k=5$ & $    0.535783 -0.121525i$ & $    0.509424 -0.11323i$ & $    0.485384 -0.105754i$  \\
						$k=6$ & $    0.535767 -0.121514i$ & $    0.509409 -0.113221i$ & $    0.485382 -0.105751i$  \\
						$k=7$ & $    0.535761 -0.121531i$ & $    0.509413 -0.113221i$ & $    0.485391 -0.105755i$  \\
						$k=8$ & $    0.535766 -0.121514i$ & $    0.509409 -0.113221i$ & $    0.485382 -0.105751i$  \\
						$k=9$ & $    0.535763 -0.121519i$ & $    0.509418 -0.113234i$ & $    0.485391 -0.105754i$  \\
						$k=10$ & $    0.535769 -0.121514i$ & $    0.509411 -0.113221i$ & $    0.485384 -0.105751i$  \\
						$k=11$ & $    0.535781 -0.121535i$ & $    0.509406 -0.113228i$ & $    0.485398 -0.105758i$  \\
						$k=12$ & $    0.535769 -0.121514i$ & $    0.509411 -0.113221i$ & $    0.485384 -0.10575i$  \\
						$k=13$ & $    0.535785 -0.121528i$ & $    0.509409 -0.113243i$ & $    0.485397 -0.105762i$  \\
						
						\hline 
					\end{tabular}%
				}
			\end{table*}
			
			\begin{table*}[!htbp]  
				\setlength{\abovecaptionskip}{0.2cm}
				\setlength{\belowcaptionskip}{0.2cm}
				\centering
				\caption{The QNM frequencies $\omega$ are presented for electromagnetic perturbations ($s=1$, $l=2$) in a black hole spacetime described by STVG with PFDM. Results are obtained via Pad\'e approximants of order $k = 1$ to $13$, with $\lambda$ fixed at $1$ and $\alpha$ varied.}
				\label{tabPADE5}
				\resizebox{\textwidth}{!}{
					\begin{tabular}{lccc}
						\hline 
						$\lambda = 1$ & $\alpha = 0.6$ & $\alpha = 0.7$ & $\alpha = 0.8$  \\
						\hline
						$k=1$ & $     0.50881 -0.112083i$ & $     0.48414 -0.104739i$ & $     0.461637 -0.098102i$  \\
						$k=2$ & $     0.500637 -0.117503i$ & $     0.476767 -0.109651i$ & $     0.454963 -0.102563i$  \\
						$k=3$ & $     0.500436 -0.118241i$ & $     0.476605 -0.110294i$ & $     0.454835 -0.103122i$  \\						
						$k=4$ & $     0.50058 -0.11836i$ & $     0.476744 -0.110401i$ & $     0.454964 -0.103216i$  \\
						$k=5$ & $     0.500575 -0.118387i$ & $     0.476734 -0.110425i$ & $     0.454956 -0.10324i$  \\
						$k=6$ & $     0.50056 -0.118374i$ & $     0.47672 -0.110412i$ & $     0.454942 -0.103229i$  \\
						$k=7$ & $     0.500568 -0.118378i$ & $     0.476727 -0.110429i$ & $     0.454941 -0.103248i$  \\
						$k=8$ & $     0.500559 -0.118374i$ & $     0.476719 -0.110412i$ & $     0.454941 -0.103229i$  \\
						$k=9$ & $     0.50057 -0.11838i$ & $     0.47673 -0.110422i$ & $     0.454948 -0.103231i$  \\
						$k=10$ & $     0.500557 -0.11838i$ & $     0.476716 -0.110416i$ & $     0.454943 -0.103229i$  \\
						$k=11$ & $     0.500578 -0.11839i$ & $     0.476721 -0.11041i$ & $     0.454958 -0.103246i$  \\
						$k=12$ & $     0.500557 -0.118379i$ & $     0.476716 -0.110416i$ & $     0.454943 -0.103228i$  \\
						$k=13$ & $     0.500578 -0.118388i$ & $     0.476721 -0.11041i$ & $      0.454958 -0.103245i$  \\
						
						\hline 
					\end{tabular}%
				}
			\end{table*}
			
			\begin{table*}[!htbp]  
				\setlength{\abovecaptionskip}{0.2cm}
				\setlength{\belowcaptionskip}{0.2cm}
				\centering
				\caption{The QNM frequencies $\omega$ are presented for axial gravitational perturbations ($s=2$, $l=2$) in a black hole spacetime described by STVG with PFDM. Results are obtained via Pad\'e approximants of order $k = 1$ to $13$, with $\lambda$ fixed at $1$ and $\alpha$ varied.}
				\label{tabPADE6}
				\resizebox{\textwidth}{!}{
					\begin{tabular}{lccc}
						\hline 
						$\lambda = 1$ & $\alpha = 0.6$ & $\alpha = 0.7$ & $\alpha = 0.8$  \\
						\hline
						$k=1$ & $      0.398303 -0.101043i$ & $      0.38096 -0.0945228i$ & $      0.365065 -0.0886197i$  \\
						$k=2$ & $      0.388391 -0.107106i$ & $      0.372039 -0.0999602i$ & $      0.357025 -0.0935209i$  \\
						$k=3$ & $      0.387495 -0.108596i$ & $      0.371394 -0.101156i$ & $      0.356573 -0.0944737i$  \\						
						$k=4$ & $      0.387213 -0.108828i$ & $      0.371299 -0.101079i$ & $      0.356557 -0.094459i$  \\
						$k=5$ & $      0.387335 -0.108457i$ & $      0.371385 -0.10105i$ & $      0.356592 -0.0944444i$  \\
						$k=6$ & $      0.386767 -0.108438i$ & $      0.371268 -0.101016i$ & $      0.356674 -0.0944i$  \\
						$k=7$ & $      0.387511 -0.108208i$ & $      0.371421 -0.101056i$ & $      0.356596 -0.0944545i$  \\
						$k=8$ & $      0.38698 -0.108432i$ & $      0.371255 -0.100975i$ & $      0.356677 -0.0943985i$  \\
						$k=9$ & $      0.38755 -0.108336i$ & $      0.371397 -0.101044i$ & $      0.356518 -0.0944237i$  \\
						$k=10$ & $      0.387035 -0.108469i$ & $      0.371245 -0.100957i$ & $      0.356678 -0.0944006i$  \\
						$k=11$ & $      0.387268 -0.108507i$ & $      0.371478 -0.10094i$ & $      0.35657 -0.0943018i$  \\
						$k=12$ & $      0.387035 -0.108469i$ & $      0.371238 -0.100959i$ & $      0.356678 -0.0944032i$  \\
						$k=13$ & $      0.387337 -0.10837i$ & $      0.371372 -0.100843i$ & $      0.356517 -0.094413i$  \\
						
						\hline 
					\end{tabular}%
				}
			\end{table*}	
	
\begin{figure*}[htbp]
		\centering
		\begin{subfigure}{0.32\textwidth}
			\includegraphics[width=\textwidth, height=5.5in, keepaspectratio]{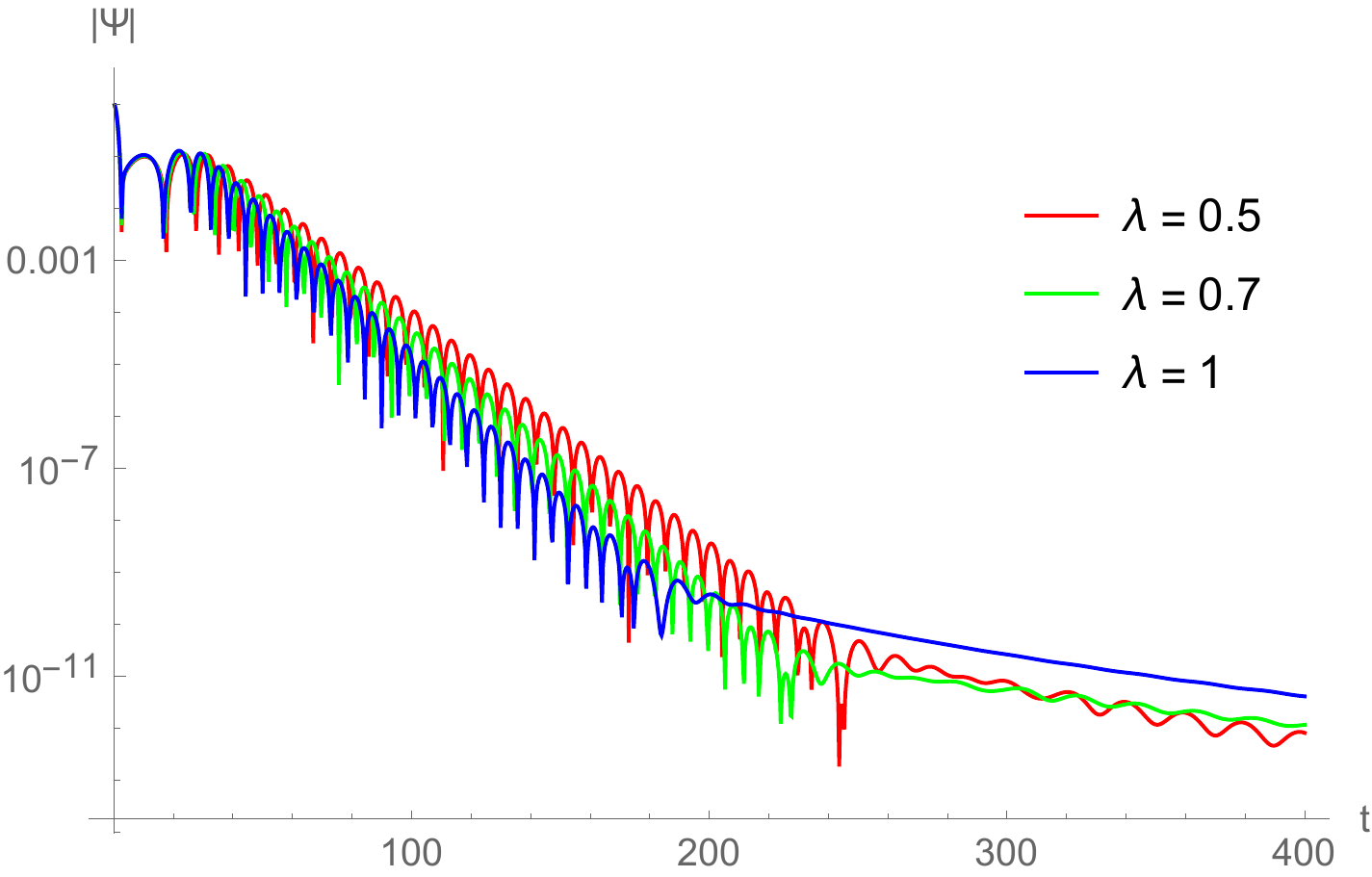}
		\end{subfigure}
		\hfill
		\begin{subfigure}{0.32\textwidth}
			\includegraphics[width=\textwidth, height=5.5in, keepaspectratio]{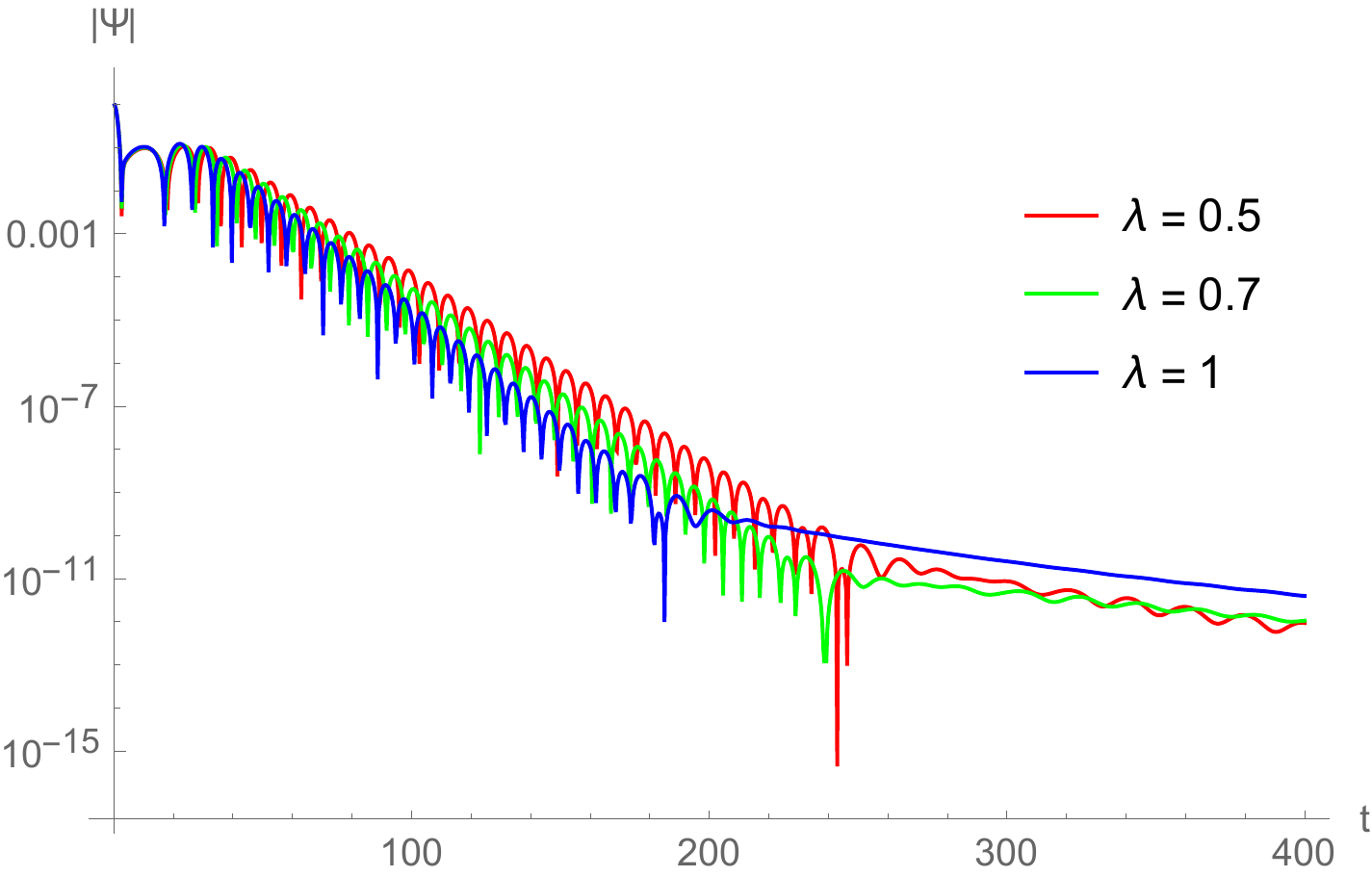}
		\end{subfigure}
		\hfill
		\begin{subfigure}{0.32\textwidth}
			\includegraphics[width=\textwidth, height=5.5in, keepaspectratio]{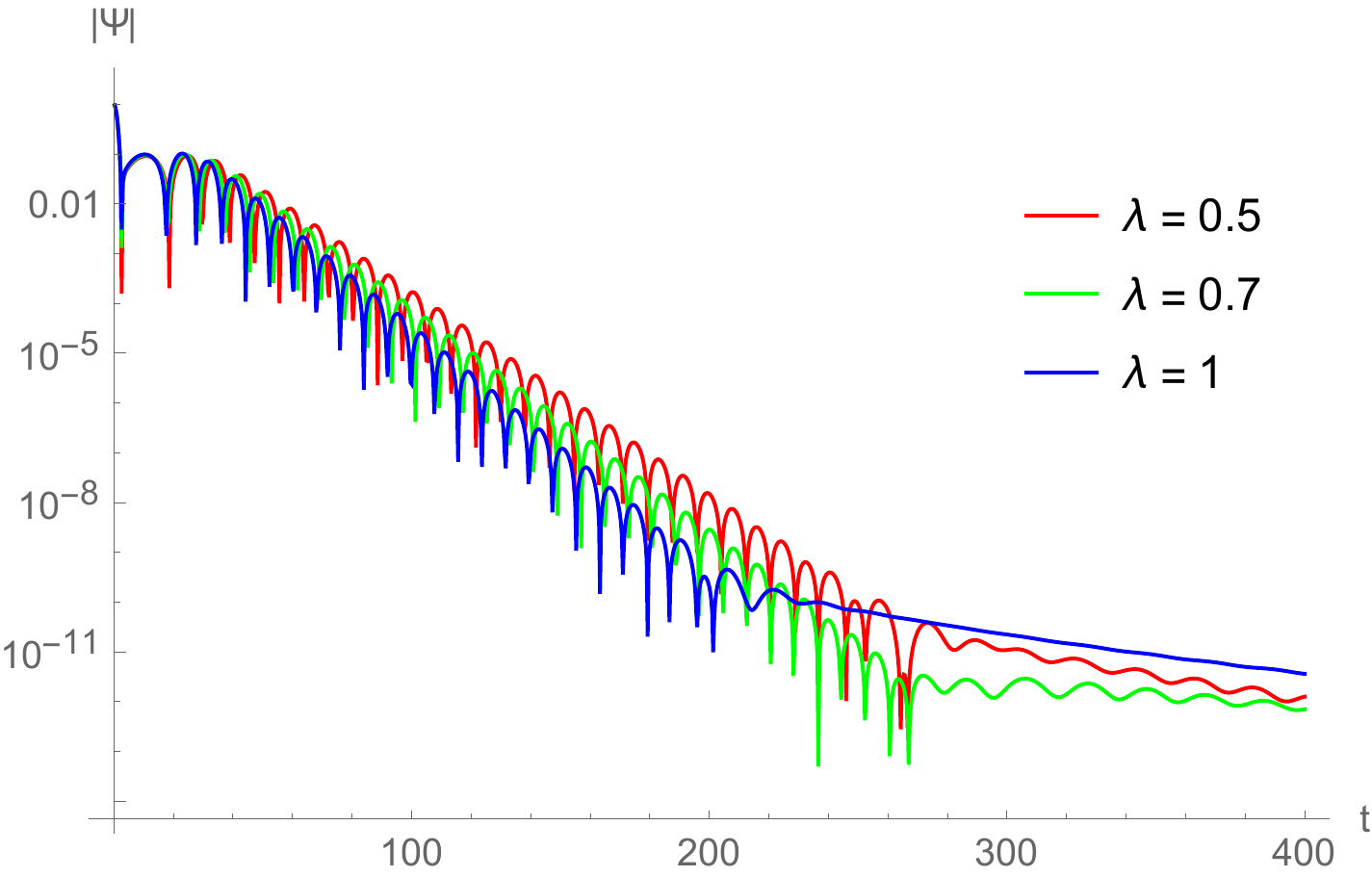}
		\end{subfigure}
		
		\vspace{0.3cm} 
		
		\begin{subfigure}{0.32\textwidth}
			\includegraphics[width=\textwidth, height=5.5in, keepaspectratio]{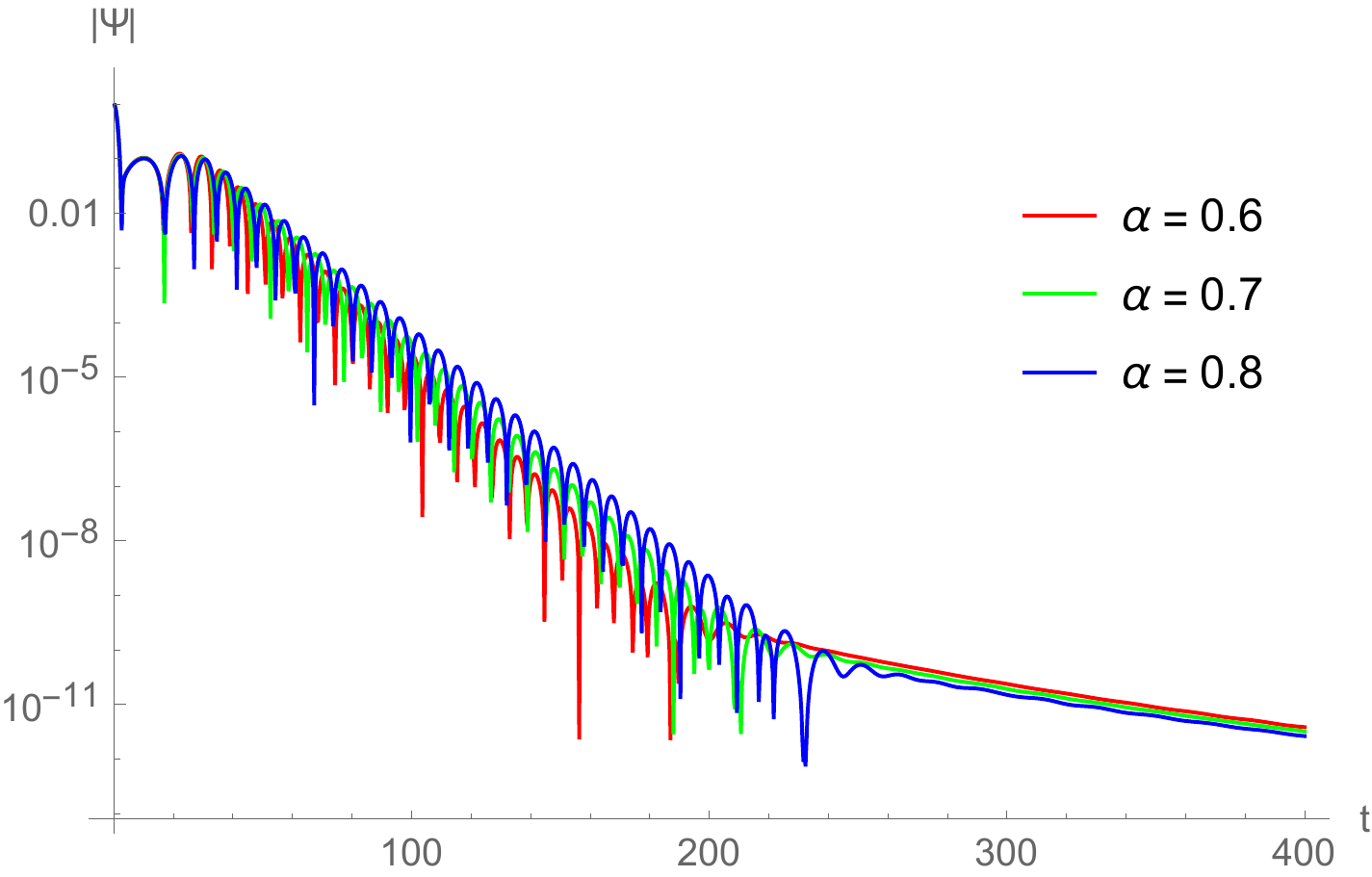}
		\end{subfigure}
		\hfill
		\begin{subfigure}{0.32\textwidth}
			\includegraphics[width=\textwidth, height=5.5in, keepaspectratio]{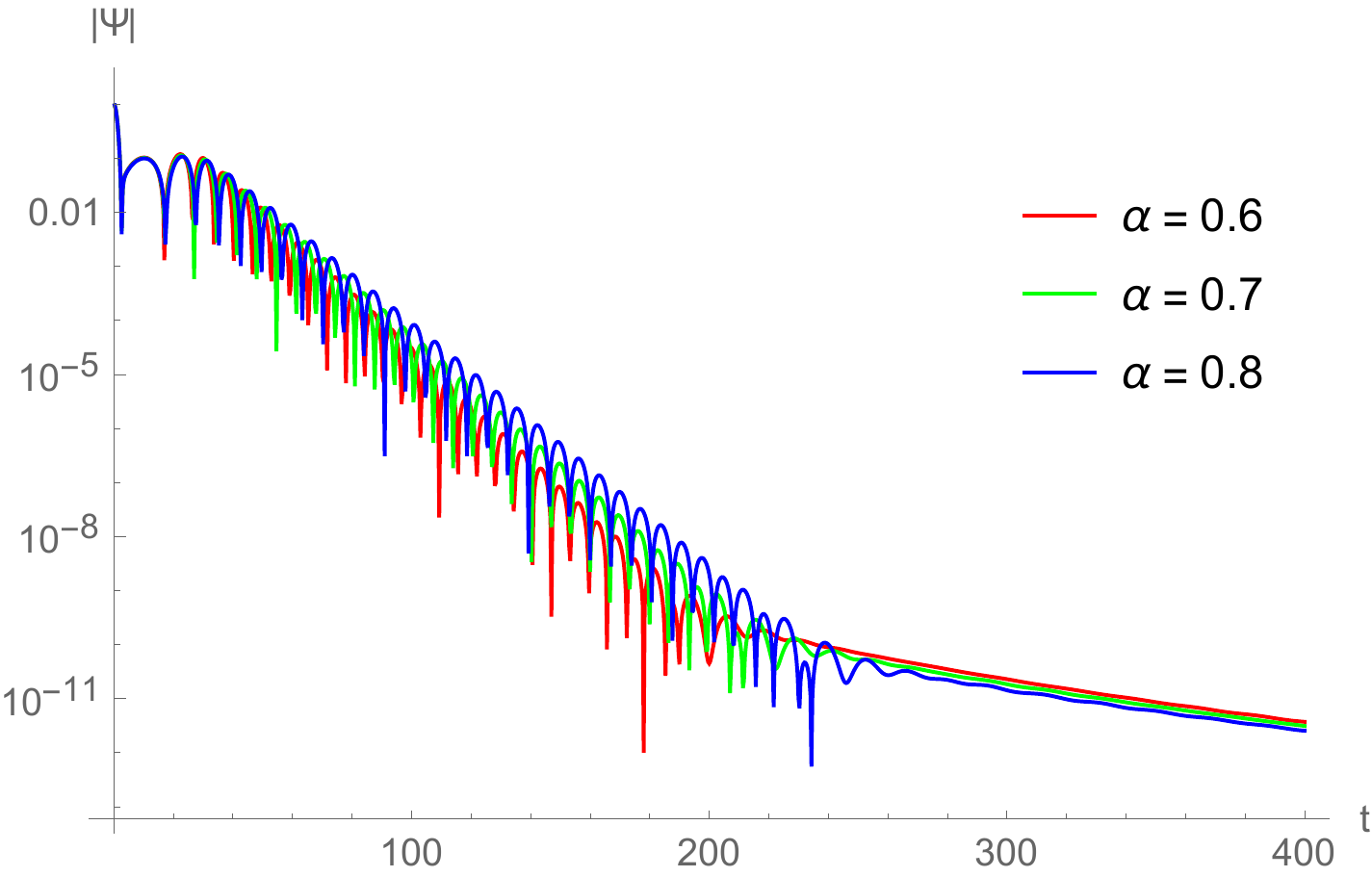}
		\end{subfigure}
		\hfill
		\begin{subfigure}{0.32\textwidth}
			\includegraphics[width=\textwidth, height=5.5in, keepaspectratio]{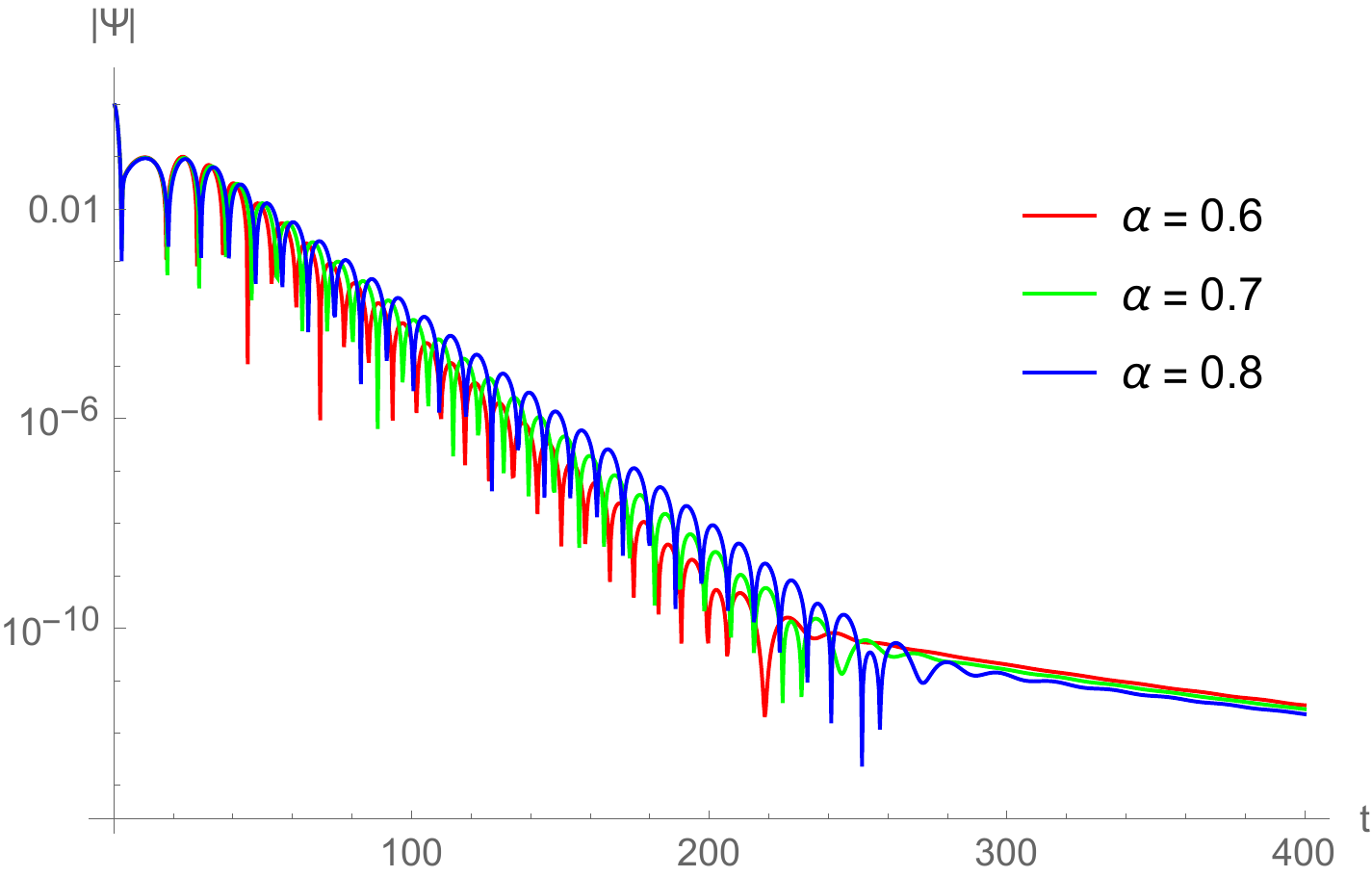}
		\end{subfigure}	
		\caption{The time-domain evolution of $l=2$ perturbations is shown for scalar ($s=0$, left), electromagnetic ($s=1$, middle), and axial gravitational ($s=2$, right) modes. The top row presents results for $\alpha = 0.55$ (fixed), whereas the bottom row displays outcomes with $\lambda = 1$ held constant.}
		\label{prony}
	\end{figure*}

	The QNM spectrum may also be obtained through numerical integration in the time domain. By employing light-cone coordinates $u = t - r_*$ and $v = t + r_*$, where $r_*$ is the tortoise coordinate, Eq. (\ref{scheq}) is cast into the form
	\begin{eqnarray}
		4\frac{\partial^{2}\psi(u,v)}{\partial u\partial v}+V(u,v)\psi(u,v)=0.\label{raodong}
	\end{eqnarray}
	A fourth-order accurate finite-difference scheme for this equation, originally introduced in Refs. \cite{Moderski:2001ru,Moderski:2001tk,Moderski:2005hf}, is given by
	\begin{align}
		\psi(N) = &\psi(W) + \psi(E) - \psi(S) \nonumber \\
		&- h^{2}\frac{V(W)\psi(W) + V(E)\psi(E)}{8} + O(h^{4}).\label{psin}
	\end{align}
	The parameter $h$ represents the step size of the numerical grid. The integration scheme utilizes the four vertices of a null rectangle: $S = (u, v)$, $E = (u, v + h)$, $W = (u + h, v)$, and $N = (u + h, v + h)$. The initial data are prescribed as a Gaussian pulse along the ingoing null surface $u = u_0$, and the field is initialized to zero on the outgoing null surface $v = v_0$:
	\begin{eqnarray}
		\psi(u=u_{0},v)=Aexp\left[-\frac{(v-v_{0})^{2}}{\sigma^{2}}\right],\psi(u,v=v_{0})=0.\label{psiu}
	\end{eqnarray}
	The time evolution of the perturbation field is simulated for scalar ($s=0$), electromagnetic ($s=1$), and axial gravitational ($s=2$) perturbations, with initial data specified by $A = 1$, $v_0 = 10$, and $\sigma = 3$. The QNM frequencies are subsequently extracted from the numerical waveforms using the Prony method \cite{Berti:2007dg,Chowdhury:2020rfj}, based on the ansatz
	\begin{eqnarray}
			\psi(t)\approx\sum_{i=1}^{p} C_{i}e^{-i\omega_{i}t}.\label{psit}
	\end{eqnarray}
	The ringdown waveforms $|\psi(t)|$ for scalar ($s=0$), electromagnetic ($s=1$), and axial gravitational ($s=2$) perturbations of a black hole in STVG with PFDM are presented in Fig. \ref{prony}. The QNM frequencies inferred from the Prony analysis (Table \ref{tabProny}) match the sixth-order WKB values to high precision across the parameter range, underscoring the reliability of both methods in STVG with dark matter.
	
		\begin{table}[htbp]
		\centering
		\caption{Fundamental QNM frequencies ($l=2$) for $s=0$, $1$, and $2$ perturbations in STVG with PFDM, obtained via the Prony method.}
		\label{tabProny}
		\adjustbox{width=0.5\textwidth}{
			\begin{tabular}{lccc}
				\hline 
				& $\omega\,(s=0)$ & $\omega\,(s=1)$ & $\omega\,(s=2)$ \\
				\hline
				\multicolumn{4}{c}{Fixed $\alpha = 0.55$} \\
				\hline
				$\lambda = 0.5$ & $    0.504413 -0.103135i$ & $   0.475342 -0.10107i$ & $   0.38137 -0.0935628i$ \\
				$\lambda = 0.7$ & $    0.533389 -0.113992i$ & $   0.500701 -0.111453i$ & $   0.395427 -0.102683i$ \\
				$\lambda = 1$ & $    0.55098 -0.125255i$ & $   0.514227 -0.122033i$ & $   0.396637 -0.111917i$ \\
				\hline
				\multicolumn{4}{c}{Fixed $\lambda = 1$} \\
				\hline
				$\alpha = 0.6$ & $    0.536733 -0.120825i$ & $   0.50137 -0.117779i$ & $  0.388024 -0.107989i$ \\
				$\alpha = 0.7$ & $    0.510232 -0.112646i$ & $   0.477411 -0.109914i$ & $  0.371871 -0.100748i$ \\
				$\alpha = 0.8$   & $    0.486086 -0.105267i$ & $   0.455534 -0.102809i$ & $   0.356998 -0.0942233i$ \\
				\hline 
			\end{tabular}
		}
	\end{table}

We now explore the connection between black hole shadow and QNMs. In the eikonal limit ($l \gg 1$), it has been established in Refs. \cite{Cardoso:2008bp,Stefanov:2010xz,Jusufi:2019ltj} that the real part of the QNM frequency is determined by the angular velocity of the unstable circular null geodesic (photon sphere), while the imaginary part is governed by the Lyapunov exponent characterizing the instability of this orbit. Specifically, the asymptotic QNM spectrum takes the form
\begin{equation}
	\omega_{S} = \omega_R - i \omega_I = \Omega \, l - i \gamma \left( n + \frac{1}{2} \right),
	\label{eikonal}
\end{equation}
where
\begin{equation}
	\Omega = \frac{\sqrt{f(r_{\mathrm{ph}})}}{r_{\mathrm{ph}}}
	\label{avelocity}
\end{equation}
is the angular velocity of photons at the photon sphere radius $r_{\mathrm{ph}}$, and
\begin{equation}
	\gamma = \sqrt{ \frac{ \big[ 2f(r_{\mathrm{ph}}) - r_{\mathrm{ph}}^{2} f''(r_{\mathrm{ph}}) \big] f(r_{\mathrm{ph}}) }{ 2 r_{\mathrm{ph}}^{2} } }
	\label{lyapunov}
\end{equation}
denotes the corresponding Lyapunov exponent.

Scalar QNM frequencies are computed for varying $l$ and $\gamma$ via Eqs. (\ref{wkbeq}) and (\ref{eikonal}), with representative results listed in Tables \ref{tabsh1}–\ref{tabsh3}. The diminishing dependence of both $\operatorname{Re}(\omega)$ and $\operatorname{Im}(\omega)$ on $l$ at large multipole numbers signals the onset of the eikonal regime, where the approximation aligns closely with the WKB results. This validates the use of the analytically simpler eikonal formula for high-$l$ modes and provides accurate input for stability analyses of compact objects.
\begin{table*}[!htbp]  
	\setlength{\abovecaptionskip}{0.2cm}
	\setlength{\belowcaptionskip}{0.2cm}
	\centering
	\caption{The quasinormal mode frequencies, obtained using the sixth-order WKB method, together with the black hole shadow radius, are listed for the case $\alpha = 0.55$ and $\lambda = 0.5$.}
	\label{tabsh1}
	\resizebox{\textwidth}{!}{
		\begin{tabular}{lcccc}
			\hline 
			$l$ & $\omega_{WKB}$ & $\omega_{S}$ & $\Delta_{R} \%$ & $\Delta_{I} \%$ \\
			\hline
			$10$ & $    2.10095 -0.103002i$ & $    2.00009 - 0.102964i$ & $    5.04277$ & $    0.0369061$ \\
			$20$ & $    4.10062 -0.102974i$ & $    4.00017 - 0.102964i$ & $    2.51114$ & $    0.00971213$ \\
			$30$ & $    6.10056 -0.102968i$ & $    6.00026 - 0.102964i$ & $    1.67159$ & $    0.00388485$ \\						
			$40$ & $    8.10057 -0.102966i$ & $    8.00034 - 0.102964i$ & $    1.25282$ & $    0.00194243$ \\
			$50$ & $    10.1006 -0.102965i$ & $    10.0004 - 0.102964i$ & $    1.00196$ & $    9.71213\times 10^{-4}$ \\
			
			\hline 
		\end{tabular}%
	}
\end{table*}

\begin{table*}[!htbp]  
	\setlength{\abovecaptionskip}{0.2cm}
	\setlength{\belowcaptionskip}{0.2cm}
	\centering
	\caption{The quasinormal mode frequencies, obtained using the sixth-order WKB method, together with the black hole shadow radius, are listed for the case $\alpha = 0.55$ and $\lambda = 0.7$.}
	\label{tabsh2}
	\resizebox{\textwidth}{!}{
		\begin{tabular}{lcccc}
			\hline 
			$l$ & $\omega_{WKB}$ & $\omega_{S}$ & $\Delta_{R} \%$ & $\Delta_{I} \%$ \\
			\hline
			$10$ & $      2.21978 -  0.113824i$ & $     2.11313 -  0.113775i$ & $     5.04702$ & $     0.0430675$ \\
			$20$ & $     4.33243 -0.113788i$ & $     4.22626 -  0.113775i$ & $     2.51215$ & $     0.0114261$ \\
			$30$ & $     6.44539 -0.113781i$ & $     6.3394 -  0.113775i$ & $     1.67192$ & $     0.00527357$ \\						
			$40$ & $     8.55844 -0.113778i$ & $     8.45253 -  0.113775i$ & $     1.253$ & $     0.00263678$ \\
			$50$ & $     10.6715 -0.113777i$ & $     10.5657 -  0.113775i$ & $     1.00135$ & $     0.00175786$ \\
			
			\hline 
		\end{tabular}%
	}
\end{table*}

\begin{table*}[!htbp]  
	\setlength{\abovecaptionskip}{0.2cm}
	\setlength{\belowcaptionskip}{0.2cm}
	\centering
	\caption{The quasinormal mode frequencies, obtained using the sixth-order WKB method, together with the black hole shadow radius, are listed for the case $\alpha = 0.55$ and $\lambda = 1$.}
	\label{tabsh3}
	\resizebox{\textwidth}{!}{
		\begin{tabular}{lcccc}
			\hline 
			$l$ & $\omega_{WKB}$ & $\omega_{S}$ & $\Delta_{R} \%$ & $\Delta_{I} \%$ \\
			\hline
			$10$ & $       2.29081 -0.124979i$ & $      2.18066 -   0.124916i$ & $      5.05122$ & $      0.0504339$ \\
			$20$ & $       4.47092 -0.124933i$ & $      4.36131 -   0.124916i$ & $      2.51324$ & $      0.0136091$ \\
			$30$ & $      6.65139 -0.124924i$ & $      6.54197 -   0.124916i$ & $      1.67258$ & $      0.0064043$ \\						
			$40$ & $      8.83195 -0.124921i$ & $      8.72262 -   0.124916i$ & $      1.25341$ & $      0.00400269$ \\
			$50$ & $      11.0125 -0.124919i$ & $      10.9033 -   0.124916i$ & $      1.00153$ & $      0.00240161$ \\
			
			\hline 
		\end{tabular}%
	}
\end{table*}
	
	\section{Conclusion} 
	\label{conclusion}
In this work, we have investigated the black hole shadow and QNM spectra in the framework of STVG coupled to PFDM, characterized by the MOG parameter $\alpha$ and the dark matter parameter $\lambda$. Our analysis reveals a clear and systematic dependence of both observables on these fundamental parameters: the shadow radius $R_{\rm sh}$ increases monotonically with $\alpha$ but decreases as $\lambda$ grows, whereas the real parts of the QNM frequencies for scalar, electromagnetic, and axial gravitational perturbations exhibit the opposite behavior—rising with $\lambda$ and falling with $\alpha$. This anti-correlated trend stems from the competing roles of modified gravity, which effectively enhances the gravitational pull at intermediate scales, and dark matter, which contributes additional attractive mass-energy near the horizon.

Most significantly, we have established an exact analytical correspondence between the shadow radius and the real part of the quasinormal frequencies in the eikonal limit ($l \gg 1$). By recognizing that both quantities are governed by the geometry of the unstable photon orbit—specifically through the critical impact parameter $b_c = r_{\rm ph}/\sqrt{f(r_{\rm ph})}$ and the photon angular velocity $\Omega = \sqrt{f(r_{\rm ph})}/r_{\rm ph}$—we derived the precise relation $\omega_R = l / b_c$, with $R_{\rm sh} \equiv b_c$ for an asymptotic observer. This prediction is robustly confirmed by independent numerical methods, including sixth-order WKB approximation, Pad\'e resummation, and time-domain integration, all of which converge to the eikonal result for large multipole numbers.

These findings underscore that the black hole shadow and gravitational ringdown are not disparate phenomena but dual observational windows onto the same underlying spacetime structure—the photon sphere. Consequently, joint measurements of shadow size (e.g., via very-long-baseline interferometry) and QNM frequencies (e.g., from future space-based gravitational-wave detectors like LISA) can provide complementary and cross-validated constraints on the parameters of modified gravity and dark matter models. Our results thus pave the way for a unified multi-messenger approach to probing strong-field gravity beyond GR.
	
	\begin{acknowledgments}
		This research was supported by the National Natural Science Foundation of China (Grant No. 12265007), and
		Guizhou Provincial Major Scientific and Technological Program XKBF (2025)010.
	\end{acknowledgments}

\end{document}